\documentclass[a4paper,11pt]{article}
\usepackage{jheppub} 
\usepackage[T1]{fontenc} 
\usepackage{undertilde,graphicx,slashed,multirow,bbm}

\newcommand{\nn}{\nonumber}
\newcommand{\bea}{\begin{eqnarray}}
\newcommand{\eea}{\end{eqnarray}}
\newcommand{\e}{\varepsilon}
\newcommand{\be}{\begin{equation}}
\newcommand{\ee}{\end{equation}}
\newcommand{\ie}{{\it i.e.}}
\newcommand{\eg}{{\it e.g.}}
\def\bsp#1\esp{\begin{split}#1\end{split}}
\def\bpm{\begin{pmatrix}}
\def\epm{\end{pmatrix}}
\def\fp{{f^\prime}}
\def\ip{{i^\prime}}
\def\Zp{{Z^\prime}}
\def\Wp{{W^\prime}}
\def\jp{{j^\prime}}
\def\kp{{k^\prime}}
\def\lp{{\ell^\prime}}
\def\lag{{\cal L}}
\def\hc{{\rm h.c.}}
\def\V{{\tilde V}}
\def\W{{\tilde W}}
\def\Q{{\tilde Q}}
\def\L{{\tilde L}}
\def\g{{\tilde g}}
\def\B{{\tilde B}}

\def\cw{{\cos\theta_W}}
\def\sw{{\sin\theta_W}}
\def\d{\mathrm{d}}
\def\tchi{{\tilde \chi}}

\preprint{CERN-PH-TH/2013-169, CUMQ/HEP 161}
\title{\Large Chargino and neutralino production at the Large Hadron Collider in left-right supersymmetric models}

\author[a]{Adam Alloul,}
\author[b]{Mariana Frank,}
\author[a,c]{Benjamin Fuks}
\author[a]{and Michel Rausch de Traubenberg}

\affiliation[a]{Institut Pluridisciplinaire Hubert Curien/D\'epartement Recherches Subatomiques,
  Universit\'e de Strasbourg/CNRS-IN2P3, 23 Rue du Loess, F-67037 Strasbourg, France}
\affiliation[b]{Department of Physics, Concordia University, 7141 Sherbrooke St.
  West, Montreal, Quebec, Canada H4B 1R6}
\affiliation[c]{Theory Division, Physics Department, CERN, CH-1211 Geneva 23,
  Switzerland}
\emailAdd{adam.alloul@iphc.cnrs.fr}
\emailAdd{mfrank@alcor.concordia.ca}
\emailAdd{fuks@cern.ch}
\emailAdd{michel.rausch@iphc.cnrs.fr}

\abstract{
We present a complete and extensive analysis of associated chargino  and neutralino production
in the framework of a supersymmetric theory augmented by left-right symmetry. This model
provides additional gaugino and higgsino states in both the neutral and charged sectors,
thus potentially enhancing new physics signals at the LHC. For a choice of
benchmark scenarios, we calculate cross sections for $7, 8$ and $14$ TeV. We  then simulate events
expected  to be produced at the LHC, and classify them according to the
number of leptons in the final state. We devise methods to reduce the background and
compare the signals with consistently simulated  events for the
Minimal Supersymmetric Standard Model. We pinpoint promising scenarios where
left-right symmetric supersymmetric signals can be distinguished both from
background and from the Minimal Supersymmetric Standard Model events.
}


\begin{document}
\maketitle
\flushbottom

 \section{Introduction}
 \label{sec:intro}

The Standard Model (SM) of electroweak interactions has been proved  enormously successful, but it leaves many important questions unanswered. It
is widely acknowledged that, from the theoretical standpoint, the SM must be
 an effective theory obtained from a more
fundamental one which is yet to be experimentally confirmed.
One of the most popular suggestions for  Beyond the Standard
Model (BSM) theories is Supersymmetry (SUSY), which introduces a new symmetry
between fundamental particles. In their simplest form, SUSY theories resolve the
gauge hierarchy and fine tuning problem, which plagues the SM, and provide a natural explanation
for the dark matter known to pervade our universe. Thus supersymmetry helps to understand the fundamental connection between particle physics and cosmology. 

A major incentive for the Large Hadron Collider (LHC) construction has been to
understand BSM processes predicted by various extensions of
the Standard Model  at high collision energies.  Since the size
of the parameter spaces is large, simplified production and decay schemes are developed to allow
 for a largely model-independent
search strategy, and benchmarks are employed to simplify the search. In supersymmetry, most of the
superpartners likely to be produced at the LHC will not be detected
as such, as they will eventually decay into the lightest supersymmetric particle
(LSP), which is stable as long as the $R$-parity  is conserved. The experimental
study of supersymmetry  hence involves  cascade decays of the
supersymmetric particles to the LSP, and the careful reconstruction of the decay
chains. The event signatures are normally characterized
by large missing transverse energy $\slashed{E}_T$, and possibly by either a high transverse
momentum  single lepton, or multilepton signals, both
with or without associated jets.

There is no evidence
so far for SUSY at the Tevatron or LEP, or   at the LHC, though the searches are far from over.
For the latter, the performance
(assuming optimal detector  efficiency)   is
significant compared to previous experiments  and one might find hints
of deviations from the Standard Model by investigating kinematical distributions associated
to signatures containing large missing  transverse energy.  Indeed, 
the  cross sections for high missing transverse energy 
 signals and for heavy particle production increase significantly when comparing
LHC $pp$ collisions at a center-of-mass energy $\sqrt{s} = 7$~TeV  or $14$ TeV with the Tevatron 
 data. One has typically a factor of  about $20$
for  $t \bar t$ pair-production,  
and up to $100$ for hypothetical particle
production  lying within the TeV range \cite{Clark:2010ka}. 

The ATLAS and CMS experiments have  been searching for supersymmetry at
the LHC during its initial run at a center-of-mass energy of
$7$ TeV and during its 2012 run at 8~TeV. Both collaborations have reported
detailed results on the limits for the SUSY partners of SM particles
\cite{ATLAS,CMS}  and the data collected
by these experiments has been able to extend searches far beyond the reach of the
Tevatron for many scenarios without, so far, any discovery or hint
for BSM results. This highlights the power of increasing the center of mass
energy.  The results are
mostly (but not exclusively) based on the assumption of minimal supergravity
and/or constrained  versions of the Minimal Supersymmetric
Standard Model (MSSM), and they set  strong direct limits on
supergravity unified models. While the model  assumption greatly simplifies
scanning the vast supersymmetric parameter space, it is important to explore the
discovery potential of superparticles in a more general context.

If a supersymmetric-like signal
is observed at the LHC, one may wonder whether the signal is indeed
supersymmetric and if the SUSY nature of the signal
is confirmed,  whether it could be explained by the minimal supersymmetric extension
of the SM or by one of the non-minimal ones. It is  thus essential 
to have clear indicators for various SUSY scenarios from phenomenology, which
could differentiate one model from another.
In supersymmetry, most event simulations have been produced assuming the underlying
model to be the MSSM, with the breaking either induced by supergravity, favoring dilepton
signatures, or by gauge interactions with a hidden sector, allowing for
diphoton  or monojet production. However, the MSSM inherits some
of the problems of the Standard Model, such as the generation of the neutrino masses
or the strong $CP$ problem.
Both can be naturally addressed within the framework of a
left-right symmetry \cite{Pati:1974yy, Mohapatra:1974hk, Mohapatra:1974gc, Senjanovic:1975rk, Mohapatra:1977mj,
Senjanovic:1978ev, Mohapatra:1979ia, Mohapatra:1995xd, Mohapatra:1996vg, Kuchimanchi:1995rp, Kuchimanchi:1993jg}.
This symmetry is favored by many
extra-dimensional models, and many gauge unification scenarios, such as $SO(10)$
 or $E_6$ \cite{Babu:1993we, Babu:1994dq, Frank:1999wz, Frank:1999ys, Mohapatra:1997sp}.
These considerations lead 
to the building of left-right supersymmetric
models (LRSUSY) based on an $SU(3)_c \times SU(2)_L \times SU(2)_R \times
U(1)_{B-L}$ gauge group \cite{Francis:1990pi,Huitu:1993uv, Huitu:1993gf}.  
Nevertheless, the open question on how these models can be distinguished from more conventional
SUSY theories as the MSSM remains.

Many existing studies addressing that question have focused on
the possibility of observing doubly-charged Higgs bosons \cite{Akeroyd:2010ip,Akeroyd:2009hb,
Ma:2009fi,Ren:2008yi,Chen:2008jh,Han:2007bk,Mukhopadhyaya:2005vf,Azuelos:2005uc,Muhlleitner:2003me,
Godfrey:2001xb,Dutta:1998bn,Huitu:1996su,Alloul:2013raa} or
doubly-charged higgsinos  \cite{Chacko:1997cm,Raidal:1998vi,Demir:2009nq,Demir:2008wt,
Frank:2007nv,Alloul:2013raa}. These new particles are predicted by a specific class
of LRSUSY models involving $SU(2)_L$ or $SU(2)_R$
triplets in the Higgs sector, responsible for the breaking of the left-right
symmetry. Of course,  the discovery of such exotic particles will be an
irrefutable  proof for an extended symmetry, 
 but will not allow to conclude about the existence of a left-right
supersymmetry. Previous
studies have  shown that the presence of extra charged
gauge bosons  associated with additional $SU(2)$
symmetries, which can thus be helicity-analyzed, 
 are indicative of left-right symmetries \cite{Frank:2010cj}.

In this work we propose to analyze the production of the fermionic
partners of the gauge and Higgs bosons, the
charginos and neutralinos, as a
signal for left-right supersymmetry. We consider
a class of left-right supersymmetric
models containing six singly-charged charginos,  the admixtures of the
supersymmetric partners of the charged gauge and Higgs bosons
of the model  and twelve neutralinos, the admixtures of the supersymmetric partners of the 
neutral gauge and Higgs bosons. By contrast, the MSSM contains only two chargino
and four neutralino states. The neutralino and chargino sectors of our LRSUSY models
contain both left-handed and right-handed gauginos,  and due
to the richness of the  spectrum,  it
is likely that several  eigenstates are light. As a consequence,
the investigation of their production and decay 
could lead to possible 
evidences for an
extended gauge structure.

Chargino-neutralino  associated production and
their subsequent decays has been extensively  searched  for by
the D$\slashed{0}$ and CDF collaborations at the Tevatron \cite{Yamaoka:2009zz},
investigating events with a $Z$ boson, decaying into $e^+e^-$, two or more jets from a
$W$ boson decay, and large missing  transverse energy.  In addition,
one of the classical  associated SUSY signature 
 consists of
the trilepton channel, where the  Standard Model background is expected to be small.
However, at the Tevatron, the chargino-neutralino trilepton signal has a low
cross section   and 
leptons are relatively hard to reconstruct as they have low transverse momentum.
At the ILC, chargino-neutralino pair production  and
 decays into the lightest neutralino (LSP) and on-shell $W$ bosons is
considered as one of the
benchmarking processes. 
Considering all-hadronic decays of the gauge bosons in the final state, one
has thus a clear 
signature of four jets with large missing energy \cite{Li:2010mq,Kafer:2009gm}. 
However, model
independent studies are difficult, and most collider results so far 
 have to be interpreted within a given
SUSY-breaking scenario, even  though phenomenological studies indicate that different
breaking mechanisms have  different implications for the spectrum of charginos and
neutralinos, which is already true  within the MSSM
\cite{Huitu:2010me}.

The LHC  being a  proton-proton machine,  it is expected to produce
squarks and gluinos copiously. Their non-observation consequently implies that
stops, squarks and gluinos are heavy if they exist. Contrary, the
charginos and neutralinos can still be rather light.
Hence, their decays into both quarks (jets) and leptons  should be
visible. Some preliminary limits on chargino and neutralino production based on MSSM
 models exist
already.   Results from the ATLAS collaboration
\cite{Aad:2012pxa,Aad:2012hba,ATLAS:2012kma,Aad:2012jja} show that
chargino masses between 110~GeV
and 340 GeV are excluded  at the 95\% confidence level in direct production of wino-like pairs
decaying into LSP via on-shell sleptons, for a 10 GeV neutralino mass.
For models with decays into intermediate degenerate sleptons,
the lightest chargino ${\tilde \chi}^+_1$ and second lightest
neutralino ${\tilde \chi}^0_2$  are even ruled out up to masses of 500
GeV.  Within the CMS experiment, final states with three leptons
in conjunction with two jets  have been used to rule out chargino and neutralino
masses between 200 and 500 GeV  for models where the branching fraction of charginos and neutralinos
into SM gauge bosons and leptons is large \cite{Chatrchyan:2012pka,Chatrchyan:2012mea,Chatrchyan:2011ff,CMS:aro}.

However,  the associated production of charginos and neutralinos in LRSUSY could yield signals that are different from those expected in the MSSM.
In this analysis we concentrate on these  and compare them with their counterparts in the MSSM in a variety of inclusive final states
involving  leptons  and missing  transverse
energy.  Since these searches require
a careful control over the SM backgrounds, 
the latter are evaluated  as well, employing state-of-the-art simulation methods.

Our work is organized as follows: in the next section (Section~\ref{sec:model}) we present a
detailed description of LRSUSY models and resolve at the same time some confusions, errors and
misconceptions in previous model versions found in the literature. We also highlight
the chargino and neutralino sectors of the theory, relevant for this work.
We then proceed with the establishment of benchmark scenarios for our
LHC simulations in Section~\ref{sec:benchmarks}. Section \ref{sec:production} is dedicated to chargino and neutralino
production and decay in terms of final states with charged leptons and missing transverse
energy. We present explicit results of an event simulation of LRSUSY signals
and compare them with Standard Model 
backgrounds (Section~\ref{sec:lrsusy}) and with their MSSM counterparts (Section~\ref{sec:mssm}).
We summarize our findings and conclude in Section~\ref{sec:conclusion}.
The Appendix contains extra details on the description of the model for
completeness.

\section{Theoretical framework: left-right symmetric supersymmetry} \label{sec:model}
In the literature, the so-called left-right supersymmetric models are based on the
$SU(3)_c \times SU(2)_L \times SU(2)_R \times U(1)_{B-L}$ gauge group.
While many versions of the model exist, we briefly describe the one used in
this paper, giving the particle content  and the
Lagrangian. We also provide the masses and
mixing matrices for the chargino and neutralino sectors, relevant for this work,
and  leave additional
details and our conventions for the Appendix.

\subsection{Particle Content}

The gauge sector of the theory is defined by
assigning one vectorial supermultiplet for each direct factor of the gauge
group, \ie, multiplets lying in the corresponding adjoint representation,
\be \bsp
  SU(3)_c \to V_3   =&\ ({\utilde{\bf 8}},{\utilde{\bf 1}},{\utilde{\bf 1}},0) \equiv 
   \big(\tilde g^a, g_\mu^a\big)\ , \\
  SU(2)_L \to V_{2L}=&\ ({\utilde{\bf 1}},{\utilde{\bf 3}},{\utilde{\bf 1}},0) \equiv 
   \big(\tilde W_L^k, W_{L\mu}^k\big)\ , \\
  SU(2)_R \to V_{2R}=&\ ({\utilde{\bf 1}},{\utilde{\bf 1}},{\utilde{\bf 3}},0) \equiv 
   \big(\tilde W_R^k, W_{R\mu}^k\big)\ , \\
  U(1)_{B-L} \to V_1=&\ ({\utilde{\bf 1}},{\utilde{\bf 1}},{\utilde{\bf 1}},0) \equiv 
   \big(\tilde{\hat B}, \hat B_\mu\big) \ .
\esp \label{eq:vectormultiplets}\ee
Here we have introduced our notations for the gauge boson and their associated
gaugino fields.

The $SU(3)_c \times SU(2)_L \times SU(2)_R \times U(1)_{B-L}$ gauge group is
broken to the Standard Model gauge group via a set of two $SU(2)_R$ Higgs
triplets $\Delta_{1R}$ and $\Delta_{2R}$ evenly charged under the $B-L$ gauge
symmetry. In addition, extra $SU(2)_L$ Higgs triplets $\Delta_{1L}$ and
$\Delta_{2L}$ are introduced to preserve parity at  higher scales.
However, the minimum of the scalar potential prefers a solution in which the
right-chiral scalar neutrinos get vacuum expectation values (vevs), breaking
$R$-parity spontaneously. Consequently, even if explicit $R$-parity
breaking is forbidden in LRSUSY models, sneutrino vevs lead to dangerous
lepton number violating operators in the superpotential. Two 
scenarios have been proposed which remedy this situation.  In Refs.\
\cite{Babu:2008ep,Frank:2011jia}, an additional singlet chiral supermultiplet ($S$) is
supplemented
to the field content of the model, leading to an $R$-parity conserving minimum of
the scalar potential after accounting for one-loop corrections. In contrast, two
extra chiral supermultiplets lying in the $({\utilde{\bf 1}},{\utilde{\bf
3}},{\utilde{\bf 1}},0)$ and $({\utilde{\bf 1}},{\utilde{\bf 1}},{\utilde{\bf
3}},0)$ representations of the LRSUSY gauge group are included in Refs.\
\cite{Aulakh:1997fq, Aulakh:1997ba}, which allow to achieve left-right symmetry breaking with
conserved $R$-parity at tree-level. We adopt the former approach, as it is a
minimal solution. The breaking of the $SU(2)_L\times U(1)_Y$ symmetry to $U(1)_{\rm em}$ is
performed by adding to the model two $SU(2)_L \times SU(2)_R$ Higgs
bidoublets $\Phi_1$ and $\Phi_2$ which are also necessary to generate
non-trivial quark mixing angles \cite{Babu:1998tm}. The field content
of the Higgs sector is thus summarized as
\be\bsp
   \qquad
      &\hspace{-1.2cm} S = \big(\utilde{\bf 1}, \utilde{\bf 1},\utilde{\bf 1},0\big) \ , \\
  (\Phi_1)^i{}_\ip = \bpm \Phi_1^0&\Phi_1^+\\ \Phi_1^-&\Phi^{\prime 0}_1{} \epm =
     \big(\utilde{\bf 1}, \utilde{\bf 2},\utilde{\bf 2}^*,0\big) \ ,
  & \qquad 
    (\Phi_2)^i{}_\ip = \bpm \Phi^{\prime 0}_2&\Phi_2^+\\ \Phi_2^-&\Phi_2{}^0 \epm  =
     \big(\utilde{\bf 1}, \utilde{\bf 2},\utilde{\bf 2}^*,0\big)\ , \\
    (\Delta_{1L})^i{}_j = 
      \bpm
        \frac{\Delta_{1L}^-}{\sqrt{2}}  & \Delta_{1L}^0\\
        \Delta_{1L}^{--} & \frac{-\Delta_{1L}^-}{\sqrt{2}} 
      \epm  = \big(\utilde{\bf 1}, \utilde{\bf 3},\utilde{\bf 1},-2\big)\ , 
 & \qquad  (\Delta_{1R})^\ip{}_\jp = 
      \bpm
        \frac{\Delta_{1R}^-}{\sqrt{2}}  & \Delta_{1R}^0\\
        \Delta_{1R}^{--} & \frac{-\Delta_{1R}^-}{\sqrt{2}} 
      \epm  = \big(\utilde{\bf 1}, \utilde{\bf 1},\utilde{\bf 3},-2\big)\ , \\
    (\Delta_{2L})^i{}_j  = 
      \bpm
        \frac{\Delta_{2L}^+}{\sqrt{2}} & \Delta_{2L}^{++}\\
        \Delta_{2L}^0 & \frac{-\Delta_{2L}^+}{\sqrt{2}} 
      \epm = \big(\utilde{\bf 1}, \utilde{\bf 3},\utilde{\bf 1},2\big) \ ,
   & \qquad  (\Delta_{2R})^\ip{}_\jp  = 
      \bpm
        \frac{\Delta_{2R}^+}{\sqrt{2}} & \Delta_{2R}^{++}\\
        \Delta_{2R}^0 & \frac{-\Delta_{2R}^+}{\sqrt{2}} 
      \epm = \big(\utilde{\bf 1}, \utilde{\bf 1},\utilde{\bf 3},2\big) \ ,
\esp\label{eq:fieldcontent1}\ee
after introducing explicitly the index structure and the
representations under the $SU(3)_c \times SU(2)_L \times SU(2)_R \times
U(1)_{B-L}$ gauge group. We recall that our conventions
regarding the $SU(2)$ indices are summarized in Appendix \ref{Sec:Model_Conv},
and the matrix
representation for the triplets, often used in the literature to build
Lagrangians, is defined by $\Delta_{a\{L,R\}} = \frac{1}{\sqrt{2}}\sigma_k
\delta_{a\{L,R\}}^k$ where $\sigma^k$ are the Pauli matrices and the
$\delta$-fields carry adjoint gauge indices $k=1,2,3$. The
electric charge $Q$ of all
fields is obtained from the well-known Gell-Mann-Nishima relation,
\be\label{eq:GellMannNishima}
  Q = T_{3L} + T_{3R} + \frac{Y_{B-L}}{2} \ ,
\ee
where $T_{3L}$, $T_{3R}$ and $Y_{B-L}$ denote the $SU(2)_L$, $SU(2)_R$ and
$U(1)_{B-L}$ quantum numbers.

In addition to the various Higgs supermultiplets described above, the chiral
sector of the theory contains left-handed ($Q_L$ and $L_L$) and right-handed
($Q_R$ and $L_R$) doublets of quark and lepton supermultiplets,
\be\bsp
    (Q_L)^{f m i  } = \bpm u_L^{f m}\\d_L^{f m} \epm =
      \big(\utilde{\bf 3}, \utilde{\bf 2},\utilde{\bf 1},\frac13\big) \ , 
 &\qquad 
    (Q_R)_{f m \ip} = \bpm u^{\mathbf c}_{R f m} & d^{\mathbf c}_{R f m} \epm=
      \big(\utilde{\bf \bar 3}, \utilde{\bf 1},\utilde{\bf 2}^*,-\frac13\big) \ ,\\
    (L_L)^{f i} = \bpm \nu_L^f \\ \ell_L^f \epm =
     \big(\utilde{\bf 1}, \utilde{\bf 2},\utilde{\bf 1},-1\big)\ ,
 & \qquad 
    (L_R)_{f \ip} = \bpm \nu_{R f}^{\mathbf c} & \ell_{R f}^{\mathbf c} \epm =
     \big(\utilde{\bf 1}, \utilde{\bf 1},\utilde{\bf 2}^*,1\big)\ ,
\esp \label{eq:fieldcontent} \ee
where the superscript ${\mathbf c}$ denotes charge conjugation, the
index $f$ is a generation index and $m$ is a color index.

\subsection{Lagrangian}
The dynamics associated to the field content presented above is described by the
Lagrangian
\be
  \lag_{\rm LRSUSY} = \lag_{\rm vector} +  \lag_{\rm chiral} + \lag_{\rm W} +
  \lag_{\rm Soft} -  V_D - V_F\ ,
\label{eq:LRSUSY} \ee
where $\lag_{\rm vector}$ and $\lag_{\rm chiral}$ contain kinetic and gauge
interaction terms associated with the vector and chiral content of the theory,
respectively, $\lag_{\rm W}$ describes the superpotential interactions between
chiral supermultiplets, $\lag_{\rm Soft}$ is the supersymmetry-breaking
Lagrangian, and the two terms $V_D$ and $V_F$ are
the so-called $D$-term and
$F$-term contributions to the scalar potential.

The gauge sector Lagrangian is fixed by gauge symmetry principles and  for
one specific vector multiplet $(\V^k, V^k_\mu)$, is
\be\bsp
 \lag_{\rm vector} = 
   &\ - \frac14 V_k^{\mu \nu} V^k_{\mu \nu} +
        \frac{i}{2}(\V^k \sigma^\mu D_\mu\overline \V_k - D_\mu \V^k\sigma^\mu
        \overline \V_k) + \ldots\ ,
\esp \ee
where the dots stand for terms included in the scalar potential contribution
$V_D$ and  $\sigma^\mu = (1,\sigma^i)$ consists of one of the possible four-vectors
built upon the Pauli matrices. In the expression above, the field strength tensor and
the covariant derivative in the adjoint representation are given by
\be 
 V_{\mu\nu}^k=\partial_\mu V_\nu^k - \partial_\nu V_\mu^k + g f_{ij}{}^k V_\mu^i V_\nu^j \ ,
 \quad
 D_\mu \V^k  = \partial_\mu\V^k   + g f_{ij}{}^k V_\mu^i \V^j \ ,
\label{eq:covderadj}\ee
where $g$ and $f_{ij}{}^k$ denote the coupling and structure constants associated to the
corresponding gauge group. The chiral Lagrangian related to one given
chiral supermultiplet $(\psi,\phi)$ is also entirely fixed by gauge invariance and
reads,
\be\bsp
  \lag_{\rm chiral} = &\ D_\mu \phi^\dag D^\mu \phi + \frac{i}{2} \Big[
    \psi \sigma^\mu D_\mu\bar\psi - D_\mu \psi \sigma^\mu \bar\psi\Big] + \Big[i\sqrt{2} 
   g \overline \V^k \cdot \bar \psi_i T_k \phi^i   + \hc\Big] + \ldots \ , 
\esp\ee
where  the dots stand for terms included in the scalar potential
contribution $V_F$. A sum over all gauge subgroups is understood, and is also included in the covariant derivative $D_\mu = \partial_\mu - i
g V^k_\mu T_k$.  An existing source of confusion in the literature refers
to the choice for the matrices $T_k$, in particular for the action of the $SU(2)_R$
symmetry. For instance, understanding the $SU(2)_L\times SU(2)_R$ structure of the Lagrangians
constructed in Refs.~\cite{Borah:2012bb,Chen:2010ss,Borah:2010kk,Dutta:1998yy,Babu:2008ep,Babu:2001se,%
Kuchimanchi:1993jg,FileviezPerez:2008sx}, which sometimes employ $SU(2)_R$ fundamental representations (in contrast to our choice)
could be not straightforward. Within those conventions, the contraction of the indices is indeed understood and therefore not trivial to get.
Furthermore, some of the analytical formulas of Refs.~\cite{Francis:1990pi,Vicente:2010wa,Rodriguez:2007pg}
contain incorrect index contractions. More precisely, for the $SU(2)_L$ gauge
  group, the generators of the Lie algebra are the Pauli matrices which act on the fields by a left action (see, \eg, the third term in the
first relation of Eq.~\eqref{eq:LR-action}), while for the $SU(2)_R$ gauge group the generators of the 
Lie algebra are minus the transpose of the Pauli matrices
and act on the fields by a right action (see, \eg, the third term in the second relation of Eq.~\eqref{eq:LR-action}). In the latter case, the
right-handed fields are indeed
in the dual of the fundamental representation of $SU(2)_R$, although equivalent to
the fundamental representation. More details are given in Appendix \ref{Sec:Model_Conv}.
 As  examples
the covariant derivatives for the left-handed and right-handed squarks $\Q_L$
and $\Q_R$, as well as  for the bidoublet of scalar Higgs fields $\Phi_1$,
read
\bea
\label{eq:LR-action}
  (D_\mu \Q_L)^{f m i} &= & 
    (\partial_\mu \Q_L)^{f m i} 
    - i g_s g^a_\mu (T_a \Q_L)^{f m i} 
    - \frac{i}{2} g_L W_{L\mu}^k (\sigma_k \Q_L)^{f m i}
    - \frac{i}{6} {\hat g}{\hat B}_\mu (\Q_L)^{f m i} \ , \nn \\
  (D_\mu \Q_R)_{f m \ip} &= &
    (\partial_\mu \Q_R)_{f m \ip} 
    \!+\! i g_s g^a_\mu (\Q_R T_a)_{f m \ip} 
    \!+\! \frac{i}2 g_R  W_{R\mu}^k (\Q_R \sigma_k)_{f m \ip} 
    \!+\! \frac{i}{6} {\hat g} {\hat B}_\mu (\Q_R)_{f m \ip} \ ,\nn \\ 
   (D_\mu \Phi_1)^i{}_\ip &=& 
     (\partial_\mu \Phi_1)^i{}_\ip 
     - \frac{i}{2} g_L W_{L\mu}^k (\sigma_k \Phi_1)^i{}_\ip
     + \frac{i}{2} g_R W_{R\mu}^k (\Phi_1 \sigma_k)^i{}_\ip\ , 
\eea 
where $g_s$, $g_L$, $g_R$ and ${\hat g}$ are the coupling constants associated to
$SU(3)_c$, $SU(2)_L$, $SU(2)_R$ and $U(1)_{B-L}$, respectively, and $T_a$ and $\sigma_k/2$
the generators of $SU(3)$ and $SU(2)$ in the fundamental representation.
Regarding the triplets of scalar Higgs fields, the covariant derivatives are given, taking the
example of the $\delta_{2L}$ and $\Delta_{2L}$ fields, by
\bea
  D_\mu \delta_{2L}^i &=& \partial_\mu\delta_{2L}^i + g_L
    \epsilon_{jk}{}^i W_{L\mu}^j \delta_{2L}^k   -i {\hat g}{\hat B}_\mu \delta_{2L}^i\ ,\\ 
  (D_\mu \Delta_{2L}){}^i{}_j &= & (\partial_\mu\Delta_{2L}){}^i{}_j -
    \frac{i}{2} g_L W_{L\mu}^k (\sigma_k \Delta_{2L})^i{}_j + \frac{i}{2} g_L
    W_{L\mu}^k (\Delta_{2L} \sigma_k)^i{}_j  -i {\hat g}{\hat B}_\mu (\Delta_{2L})^i{}_j \ ,
\nonumber\eea
where $\epsilon$ is the rank-three antisymmetric tensor related to the adjoint
representation of $SU(2)$. In the first line of the equation above, we
have used the common form for fields lying in the adjoint representation
and in the second line, the matrix representation for $SU(2)$ triplets
introduced in Eq.\ \eqref{eq:fieldcontent1}.

The most general superpotential describing the interactions among the model chiral
supermultiplets is
\bea
   W(\phi) &=&\nonumber
   (\Q_L)^{m i} y_Q^1 (\hat \Phi_1)_i{}^\ip (\Q_R)_{m \ip} +
   (\Q_L)^{m i} y_Q^2 (\hat \Phi_2)_i{}^\ip (\Q_R)_{m \ip} +
   (\L_L)^i y_L^1 (\hat \Phi_1)_i{}^\ip (\L_R)_\ip \\
&&\nonumber
   + (\L_L)^i y_L^2 (\hat \Phi_2)_i{}^\ip (\L_R)_\ip + 
   (\hat \L_L)_i y_L^3 (\Delta_{2L})^i{}_j (\L_L)^j +
   (\L_R)_\ip y_L^4 (\Delta_{1R})^\ip{}_\jp (\hat\L_R)^\jp \\
&&\nonumber
   + \Big( \mu_L + \lambda_L S \Big)\Delta_{1L} \cdot \hat \Delta_{2L}
   + \Big( \mu_R + \lambda_R S \Big)\Delta_{1R} \cdot \hat \Delta_{2R} 
   + \Big( \mu_3 + \lambda_3 S \Big)\Phi_1 \cdot \hat{\Phi}_2 \\
&&
   + \frac13 \lambda_s S^3 + \mu_s S^2 + \xi_F S\ ,
\label{eq:Wtrip}\eea
where squark and slepton flavor indices are understood
(the Yukawa couplings $y_Q$ and $y_L$ are $3\times 3$ matrices in flavor space).
This superpotential is expressed in terms of the scalar degrees of freedom of the
field content of the theory, \ie, squarks and sleptons $\Q_L$, $\Q_R$, $\L_L$
and $\L_R$, the Higgs fields $\Phi_i$ and $\Delta_{a{\{L,R\}}}$, and the singlet
field $S$. We have also introduced the hatted ( $\hat {}$ ) quantities
\be\bsp
 &\ (\hat \L_L)_i = \e_{ij} (\L_L)^j \ , \quad
  (\hat \L_R)^\ip = \e^{\ip\jp} (\L_R)_\jp  \ , \quad
  (\hat  \Phi_{1,2})_i{}^\ip = \e^{\ip\jp} \e_{ij} ( \Phi_{1,2})^j{}_\jp \ ,  \\ 
 &\ (\hat \Delta_{2L})_i{}^j = \e_{ik} \e^{j\ell} (\Delta_{2L})^k{}_\ell \
, \quad
   (\hat \Delta_{2R})_\ip{}^\jp = \e_{\ip\kp} \e^{\jp\lp} (\Delta_{2R})^\kp{}_\lp \ , 
\esp\ee
and the associated invariant products
\be\bsp
 &\ \Delta_{1L} \cdot \hat \Delta_{2L} \equiv 
    \text{Tr}\Big(\Delta^t_{1L} \hat \Delta_{2L}\Big) =
    (\Delta_{1L})^i{}_j \ (\hat \Delta_{2L})_i{}^j \ , \\ 
&\  \Delta_{1R} \cdot \hat \Delta_{2R} \equiv 
    \text{Tr}\Big(\Delta^t_{1R} \hat \Delta_{2R}\Big) =
    (\Delta_{1R})^\ip{}_\jp \ (\hat \Delta_{2R})_\ip{}^\jp \\
 &\  \Phi_1 \cdot \hat \Phi_2 \ \equiv \ \text{Tr}( \Phi_1^t \hat  \Phi_2) = ( \Phi_1)^i{}_{\ip} (\hat  \Phi_2)_i{}^\ip
    \ ,
\esp
\ee
as well as the Yukawa couplings  $\lambda$, the bilinear
supersymmetric mass terms $\mu$ and the linear $\xi$-term. We recall that the
conventions for the $SU(2)$ invariant tensors $\e_{ij}$ and $\e^{ij}$ are indicated in the Appendix.
Left-right symmetry requires all ${y}^{1,2}_Q, y^{1,2}_L$ matrices to be
Hermitian in the generation space  and  ${y}^{3,4}_{L}$ matrices to be
symmetric.
 The superpotential can however be simplified by neglecting all the
three $\mu$-terms mixing 
the different Higgs fields, keeping only the effective bilinear terms
dynamically generated by the vev of the singlet field. This limit, motivated by,
\eg, a discrete $\mathbbm{Z}_3$ symmetry of the superpotential, can naturally
explain both the strong and SUSY $CP$ problems \cite{Babu:2008ep}.

 The  soft-SUSY breaking Lagrangian of Eq.\ \eqref{eq:LRSUSY} is given by
\be\bsp
 \lag_{\rm Soft} = &\ 
   - \frac12 \bigg[ M_1 \tilde{\hat{B}}\cdot \tilde{\hat{B}} + M_{2L} \W_L^k \cdot \W_{Lk} + M_{2R} \W_R^k \cdot
      \W_{Rk} + M_3 \g^a \cdot \g_a + \hc \bigg] \\ 
    &\ - \Q_L^\dag m_{Q_L}^2 \Q_L 
       - \Q_R m_{Q_R}^2 \Q_R^\dag  
       - \L_L^\dag m_{L_L}^2 \L_L
       - \L_R m_{L_R}^2 \L_R^\dag \\
  &\
       - (m_\Phi^2)^{\fp f} {\rm Tr} (\Phi_f^\dag \Phi_\fp)
     - m_{\Delta_{1L}}^2 {\rm Tr} (\Delta_{1L}^\dag \Delta_{1L})
       - m_{\Delta_{2L}}^2 {\rm Tr}(\Delta_{2L}^\dag \Delta_{2L}) 
\\ &\
       - m_{\Delta_{1R}}^2 {\rm Tr} (\Delta_{1R}^\dag \Delta_{1R}) 
       - m_{\Delta_{2R}}^2 {\rm Tr} (\Delta_{2R}^\dag \Delta_{2R})
       - m_s^2 S^\dag S
\\ &\
  - \bigg[ 
        \Q_L T_Q^1 \hat{\Phi}_1 \Q_R 
      + \Q_L T_Q^2 \hat{\Phi}_2 \Q_R
      + \L_L T_L^1 \hat{\Phi}_1 \L_R
      + \L_L T_L^2 \hat{\Phi}_2 \L_R
      + \hat{\L}_L T_L^3 \Delta_{2L} \L_L
\\ &\quad
      + \L_R T_L^4  \Delta_{1R} \hat{\L}_R 
      + \Big( B_L+  T_L S\Big)\Delta_{1L} \cdot \hat{\Delta}_{2L} 
      + \Big( B_R+  T_R S\Big)\Delta_{1R} \cdot \hat{\Delta}_{2R} 
\\&\quad
      + \Big( B_3+  T_3 S\Big)\Phi_1 \cdot \hat{\Phi}_2 +
    \frac13 T_s S^3 + B_s S^2 + \xi_s S  + \hc \bigg]  \ . 
\esp \ee
where all the indices but the bidoublet flavor ones are understood.
In this last expression, the first line provides mass terms for the gaugino
fields, the three next lines mass terms for all
the scalar fields and the other lines are derived from the form of the
superpotential,
the trilinear couplings  $T_Q$ and $T_L$ being $3\times 3$ matrices in flavor space.
Finally, the $F$-term and $D$-term contributions to the scalar potential $V_F$
and $V_D$ are obtained after solving the equations of motion for the auxiliary fields associated
with each supermultiplet,
\be
  V_F = \frac{\partial W(\phi)}{\partial \phi^i} \frac{\partial
     W^\dag(\phi^\dag)}{\partial \phi^\dag_i} 
  \quad \text{and}\quad
  V_D = \frac12 g^2 (\phi^\dag T^k \phi)(\phi^\dag T_k \phi)  \ ,
\label{eq:LFD}\ee
where sums over all the direct factors of the gauge group and  the chiral
content of the theory are understood.

The gauge symmetry is spontaneously broken in two steps, the $SU(2)_R\times
U(1)_{B-L}$ gauge group being first broken to the electroweak gauge group which is
subsequently broken to the electromagnetic group $U(1)_{\rm em}$. At the minimum of the scalar potential, the
neutral components of the Higgs fields obtain non-zero vevs,
\be\bsp
   & \langle S \rangle = \frac{v_s}{\sqrt{2}} e^{i
    \alpha_s}\ ,
  \quad
  \langle \Phi_1 \rangle =  \bpm 
     \quad \frac{v_1}{\sqrt{2}} \quad & 0 \\ 
      0 & \frac{v_1^\prime}{\sqrt{2}}  e^{i \alpha_1} \epm \ , 
  \quad 
  \langle \Phi_2 \rangle = \bpm 
     \quad \frac{v_2^\prime }{\sqrt{2}} e^{i \alpha_2} \quad & 0\\ 
     0 &\frac{v_2}{\sqrt{2}} \epm \ , \\
  &
  \langle \Delta_{1L} \rangle =  \bpm 0 & \frac{v_{1L}}{\sqrt{2}} \\ 0 & 0 \epm , \ 
  \langle \Delta_{1R} \rangle =  \bpm 0 & \frac{v_{1R}}{\sqrt{2}} \\ 0 & 0 \epm , \
  \langle \Delta_{2L} \rangle =  \bpm 0 & 0 \\ \frac{v_{2L}}{\sqrt{2}} & 0 \epm , \
  \langle \Delta_{2R} \rangle =  \bpm 0 & 0 \\ \frac{v_{2R}}{\sqrt{2}} & 0 \epm .
\esp\ee
Keeping the number of independent complex phases  minimum, the vevs $v_{iL}$,
$v_{iR}$, $v_1$, $v_2$, $v^\prime_1$, $v^\prime_2$ and $v_s$ can be chosen real and
non-negative whilst the only complex phases which cannot be rotated away by
means of suitable gauge transformations  and field redefinitions are
denoted by $\alpha_1$, $\alpha_2$ and $\alpha_s$.
 This rather large
number of degrees of freedom  can be reduced  by the strong constraints existing on the different vevs. Although 
in the
supersymmetric limit, the vev of the singlet field is vanishing, it becomes of
the order of the supersymmetry-breaking scale after SUSY-breaking. 
Since  $v_{1R}$ and $v_{2R}$
are related to the masses of the $SU(2)_R$ gauge bosons and to the Majorana masses
of the right-handed neutrinos, they must be larger than the other vevs related
to the SM-like particle masses. In addition, small left-handed neutrino Majorana
masses require that the vevs of the $SU(2)_L$ Higgs triplets, $v_{1L}$ and
$v_{2L}$, are negligibly small. Finally, as it is shown in Appendix
\ref{app:ewsb}, the possibly $CP$-violating $W^\pm_L-W^\pm_R$ mixing is dictated
by the products $v_1 v_1^\prime e^{i \alpha_1}$ and $v_2 v_2^\prime e^{i
\alpha_2}$ which is constrained to be small by $K^0-{\bar K}^0$ mixing data.
Hence, we assume the hierarchy
\be \label{eq:vevhier}
  v_s \gg v_{1R}, v_{2R} \gg  v_2, v_1\gg v_1^\prime  = v_2^\prime =
   v_{1L} = v_{2L} \approx 0 \qquad\text{and}\qquad
 \alpha_1 = \alpha_2 \approx 0 \ .
\ee

\subsection{Charginos and neutralinos}
In the fermionic sector, all the partners of the gauge and Higgs bosons with the same
quantum numbers (electric charge and color representation) mix after breaking the
electroweak symmetry  to electromagnetism. The model contains twelve
neutralinos, the admixtures of the neutral superpartners. Their symmetric mass
matrix, expressed in the $(i
\W_L^3, i \W_R^3, i \B, \tilde \Phi_2^{\prime 0}, \tilde \Phi_2^0,
\tilde \Phi_1^0, 
\tilde \Phi_1^{\prime 0},$ $\tilde{\Delta}_{2L}^0, \tilde{\Delta}_{2R}^0,
\tilde{\Delta}_{1L}^0, \tilde{\Delta}_{1R}^0, \tilde S)$ basis, reads
\renewcommand{\arraystretch}{1.3}
\bea
&& M_{\chi^0}=\\
&&\nonumber
 \small{ \! \!\!\left(\begin{array}{c c c c c c c c c c c c}
      M_{2L} & 0 & 0 & \frac{g_L \tilde v_2^\prime}{2} & -\frac{g_L v_2 }{2} & 
        \frac{g_L v_1 }{2} & -\frac{g_L \tilde v_1^\prime }{2} & -g_L v_{2L} & 0 &
        g_L v_{1L} & 0 & 0\\ 
      0 & M_{2R} & 0 & -\frac{g_R \tilde v_2^\prime}{2}  & \frac{g_R v_2}{2} & 
        -\frac{g_R  v_1 }{2}& \frac{g_R \tilde v_1^\prime}{2}  & 0 & -g_R v_{2R} &  
        0  & g_R v_{1R} & 0\\
      0 & 0 & M_1 & 0 & 0 & 0 & 0 & {\hat g} v_{2L} & {\hat g} v_{2R} & 
        -{\hat g} v_{1L}& -{\hat g} v_{1R} &0 \\
      \frac{g_L \tilde v_2^\prime}{2}  & -\frac{g_R \tilde v_2^\prime }{2} & 0 & 0
        & 0 & 0 & -\tilde\mu_3 & 0 & 0 & 0 & 0 & -\frac{\lambda_3 \tilde
        v_1^\prime}{\sqrt{2}} \\ 
      -\frac{g_L  v_2 }{2}&  \frac{g_R v_2}{2}  & 0 & 0 & 0 & 
         -\tilde\mu_3 & 0 & 0 & 0 &0 &0&
     -\frac{\lambda_3 v_1}{\sqrt{2}}\\
      \frac{g_L v_1}{2}  & -\frac{g_R  v_1}{2} & 0 & 0 &-\tilde\mu_3&0&0&0&0&0&0
& -\frac{\lambda_3 v_2}{\sqrt{2}}\\
      -\frac{g_L  \tilde{v}_1^\prime}{2}  & \frac{g_R \tilde v^\prime_1}{2}  & 0 &
      -\tilde\mu_3  &0& 0&0&0&0&0&0 &  -\frac{\lambda_3 \tilde v_2^\prime}{\sqrt{2}} \\
      -g_L v_{2L} & 0 & {\hat g} v_{2L}&0&0&0&0&0&0&\tilde\mu_L&0 & \frac{\lambda_L v_{1L}}{\sqrt{2}}
\\
      0&-g_R v_{2R} & {\hat g} v_{2R}&0&0&0&0&0&0&0&\tilde\mu_R &\frac{\lambda_R v_{1R}}{\sqrt{2}}  \\
      g_L v_{1L} & 0 & -{\hat g}v_{1L}&0&0&0&0&\tilde\mu_L&0&0&0&  \frac{\lambda_L v_{2L}}{\sqrt{2}}
\\
      0&g_R v_{1R} & -{\hat g} v_{1R}&0&0&0&0&0&\tilde \mu_R &0&0& \frac{\lambda_R v_{2R}}{\sqrt{2}} \\
      0& 0 & 0 & -\frac{\lambda_3 \tilde v_1^\prime}{\sqrt{2}} &
     -\frac{\lambda_3 v_1}{\sqrt{2}}& -\frac{\lambda_3 v_2}{\sqrt{2}}&
     -\frac{\lambda_3 \tilde v_2^\prime}{\sqrt{2}} & \frac{\lambda_L v_{1L}}{\sqrt{2}}
     & \frac{\lambda_R v_{1R}}{\sqrt{2}} & \frac{\lambda_L v_{2L}}{\sqrt{2}}
    & \frac{\lambda_R v_{2R}}{\sqrt{2}} & 2\tilde \mu_s
    \end{array}\right) \ ,
}\label{eq:mneu}\eea
with
\be
  \tilde v_i^\prime = v_i^\prime e^{i \alpha_i} \qquad\text{and}\qquad
   \tilde\mu_{L,R,3,S} = \mu_{\{L,R,3,S\}} + \frac{1}{\sqrt{2}}
\lambda_{\{L,R,3,S\}} v_s e^{i \alpha_s} \ . 
\ee
This matrix is diagonalized
through a unitary matrix $N$ which relates the twelve physical (two-component)
neutralinos $\chi^0_i$ to the interaction eigenstates,
\be\bsp 
 &\ ( \chi^0_1 \ \ \chi^0_2 \ \ \chi^0_3 \ \ \chi^0_4 \ \ \chi^0_5 \ \ \chi^0_6 \
    \ \chi^0_7 \ \ \chi^0_8 \ \ \chi^0_9 \ \ \chi^0_{10} \ \ \chi^0_{11} \ \
    \chi^0_{12})^t =\\
 &\quad N
  (i \W_L^3\ \  i \W_R^3 \ \ i \B \ \ \ \tilde\Phi_2^{\prime 0} \ \ \ \tilde\Phi_2^0 \ \ \
\tilde\Phi_1^0\ \
    \ \tilde \Phi_1^{\prime 0}\ \ \tilde{\Delta}_{2L}^0 \ \ \tilde{\Delta}_{2R}^0 \ \
    \tilde{\Delta}_{1L}^0\ \ \tilde{\Delta}_{1R}^0\ \ \tilde{S} )^t  \ .
\esp\ee

Turning to the charged sector, the model contains six singly-charged charginos, the charged superpartners of the
gauge and Higgs bosons. The associated mass matrix, given in the $(i \W_L^+, i
\W_R^+, \ \tilde\Phi_2^+, \ \tilde\Phi_1^+, \tilde{\Delta}_{2L}^+, \tilde{\Delta}_{2R}^+)$ and $(i
\W_L^-, i \W_R^-, \ \tilde\Phi_2^-, \ \tilde\Phi_1^-, \tilde{\Delta}_{1L}^-, \tilde{\Delta}_{1R}^-)$ bases by
\be\bsp
M_{\chi^\pm}= \left(\begin{array}{c c c c c c}
      M_{2L} & 0 & \frac{g_L}{\sqrt{2}}\tilde  v_2^\prime & \frac{g_L}{\sqrt{2}} v_1 & 
       -g_L v_{1L} & 0\\ 
      0 & M_{2R} & -\frac{g_R}{\sqrt{2}} v_2  & 
        - \frac{g_R}{\sqrt{2}} \tilde v_1^\prime & 0 & -g_R v_{1R} \\
      \frac{g_L}{\sqrt{2}} v_2 & -\frac{g_R}{\sqrt{2}}  \tilde v_2^\prime & 0 &
        \tilde \mu_3 & 0 & 0 \\
      \frac{g_L}{\sqrt{2}} \tilde  v^\prime_1 & -\frac{g_R}{\sqrt{2}} v_1 &
       \tilde\mu_3  
         &0& 0&0\\
      g_L v_{2L} & 0 &0&0&\tilde\mu_L&0\\
      0&g_R v_{2R} & 0&0&0&\tilde\mu_R\\
    \end{array}\right) \ ,
\esp\label{eq:mcharg}\ee
is diagonalized through two unitary  rotations
$U$ and $V$ relating the
interaction eigenstates to the physical (two-component) charginos eigenstates
$\chi^\pm_i$,
\be\bsp
  ( \chi^+_1 \ \ \chi^+_2 \ \ \chi^+_3 \ \ \chi^+_4 \ \ \chi^+_5 \ \ \chi^+_6)^t
    =&\ V (i \W_L^+ \ \ i \W_R^+\ \ \ \tilde\Phi_2^+ \ \ \
\tilde\Phi_1^+\ \ \tilde{\Delta}_{2L}^+\ \
    \tilde{\Delta}_{2R}^+)^t\ , \\
  ( \chi^-_1 \ \ \chi^-_2 \ \ \chi^-_3 \ \ \chi^-_4 \ \ \chi^-_5 \ \ \chi^-_6)^t
    =&\ U (i \W_L^- \ \ i \W_R^-\ \ \ \tilde\Phi_2^- \ \ \ \tilde\Phi_1^-\ \ \tilde{\Delta}_{1L}^-\ \
    \tilde{\Delta}_{1R}^-)^t \ . 
\esp\ee

LRSUSY models also contain four doubly-charged charginos, the fermionic
partners of the doubly-charged Higgs bosons. We include them for completeness,
although their phenomenology at colliders has been widely studied  in the past
\cite{Chacko:1997cm,Raidal:1998vi,Demir:2009nq,Demir:2008wt,
Frank:2007nv}. Their mass matrix, which is already diagonal and does not
need  to be  further rotated, is expressed in the $(\chi^{++}_1, \chi^{++}_2) =
(\tilde{\Delta}_{2L}^{++}, \tilde{\Delta}_{2R}^{++})$ and
$(\chi^{--}_1, \chi^{--}_2) = (\tilde{\Delta}_{1L}^{--},
\tilde{\Delta}_{1R}^{--})$ bases as
\be\bsp
M_{\chi^{\pm \pm}}= \left(\begin{array}{c c}
    \tilde\mu_L&0\\
    0&\tilde\mu_R\\
 \end{array}\right) \ .
\esp\ee

Before moving on, we recall that the (commonly used) four-component
representations for neutralinos and charginos are defined as, 
\be 
    \tilde\chi^0_i = \bpm \chi^0_i \\ \bar \chi^{0i} \epm \ , \quad 
    \tilde\chi^\pm_i= \bpm \chi^\pm_i \\ \bar \chi^\mp_i \epm \ , \quad
    \tilde\chi^{\pm\pm}_i = \bpm \chi^{\pm\pm}_i \\ \bar \chi_i^{\mp\mp} \epm \ .
\ee

\section{Benchmark scenarios}
\label{sec:benchmarks}
In this Section, we construct a set of several benchmark scenarios for
neutralino and chargino phenomenology at hadron colliders in the context
of LRSUSY models. Due to the large
number of free parameters in the theory, we consider a restricted version of the
model presented in the previous Section. First, the superpotential of Eq.\
\eqref{eq:Wtrip} is simplified by assuming a discrete $\mathbbm{Z}_3$ symmetry
where each scalar field transforms as
\be
   \phi \to e^{\frac{2 \pi i}{3}} \phi \ .
\ee
Consequently, all bilinear and linear terms are forbidden. However,
the $\mathbbm{Z}_3$ symmetry is spontaneously broken by the vevs of the Higgs
fields, which leads to well-known domain-wall issues \cite{Zeldovich:1974uw,
Vilenkin:1984ib}. These problems can be avoided
by including higher-dimensional, non-renormalizable,
Planck-scale suppressed operators in the superpotential
so that at the LHC energy range, the superpotential of Eq.\ \eqref{eq:Wtrip} is left unchanged.

Second, the hierarchy among the vacuum expectation values of Eq.\
\eqref{eq:vevhier} allows to simplify the number of degrees of freedom related
to the Higgs sector. Neglecting very small vevs, we have
further assumed the singlet vev to be real ($\alpha_s=0$)
and far above the SUSY-breaking scale, which is possible  with not too
large $\lambda$-parameters in the superpotential.

Furthermore, the parameters of the electroweak sector are not all
independent. Hence,  at tree-level, the three gauge coupling constants $g_L$, $g_R$ and
${\hat g}$ are related to the $Z$- and $W$-boson mass $m_Z$ and $m_W$ and  to
the electroweak coupling constant at the $Z$-pole,
$\alpha$. Assuming the left-right symmetry of the coupling constants to survive
at the weak scale, one imposes in addition $g_L = g_R${\footnote{We choose the formal left-right symmetric condition $g_L=g_R$ for simplicity. In principle, the condition $g_R>g_L \tan \theta_W$
has to be satisfied, otherwise the couplings $Z_Rf {\bar f}$ become non-perturbative. Also, if $g_R>g_L$, right-handed currents would dominate over the left-handed ones. Thus choosing $g_R=g_L$ is not particularly restrictive.}. Hence, 
on the basis of the relations
presented in Appendix~\ref{app:ewsb}, we have
\be
   \cos\theta_W = \frac{m_W}{m_Z} \ , \
   e = \sqrt{4 \pi \alpha} \ , \
   g_R = g_L = \frac{e}{\sin\theta_W}\  ,  \
   {\hat g} = \frac{e}{\sqrt{\cos 2\theta_W}} \ , \
   v = \frac{2 \cos\theta_W m_Z}{g_L} \ , 
\ee
where the electroweak inputs are  $m_Z = 91.1876$ GeV, $m_W =80.399$ GeV
and $\alpha(m_Z)^{-1} = 127.9$ \cite{Beringer:2012zz}.
The neutralino mass matrix of Eq.\ \eqref{eq:mneu} is then related, at tree-level,
to the eleven free parameters,
\be
  M_1\ , \ \ M_{2L}\ , \quad M_{2R}\ , \ \ 
  v_R \ , \ \ \tan\beta=\frac{v_2}{v_1} \ , \ \
  \tan\tilde\beta=\frac{v_{2R}}{v_{1R}}\ ,  \ \  v_s \ , \ \
  \lambda_L \ , \ \
  \lambda_R \ , \ \
  \lambda_s \ , \ \
  \lambda_3  \ .
\label{eq:params}\ee
We assume  in addition that $v_s$
 is of order ${\cal O}(100\text{ TeV})$ and $v_R$, defined in Appendix~\ref{app:ewsb},
of the order of the TeV scale.

The large values of the right-handed vevs $v_{1R}$ and
$v_{2R}$, together with the one of the singlet vev $v_s$, shift all the
higgsino fields to a higher scale. Therefore, we are left to consider the three
lighter neutralino states and the two lighter chargino states, 
which are admixtures of the bino and the two wino gauge eigenstates\footnote{This choice
highlights the gauge structure of LRSUSY and sets it apart from other models.}.
According to the different possible hierarchies between the three
soft gaugino masses $M_1$, $M_{2L}$ and $M_{2R}$, one can
in principle envisage different mixing scenarios, as it will be shown below.

We now turn to the sfermion sector. 
Inspired by some organizing principle based on unification at high energy, we
choose to decouple squarks and gluinos. As in the MSSM, 
renormalization group running down to the electroweak scale 
shifts squark and gluino masses to a significantly  higher scale compared to the 
slepton and sneutrino  ones due to strong
contributions to the various beta functions of the soft parameters. Under this
assumption, the sfermion sector is entirely defined
by supplementing to the parameters presented in Eq.~\eqref{eq:params} the soft
masses related to left-handed and right-handed sleptons and sneutrinos. Taking
them flavor-universal, one has  two new free parameters,
\be
  m_{\tilde L_L} \qquad \text{and} \qquad
  m_{\tilde L_R} \ . 
\ee
Furthermore, slepton mixing, proportional to the lepton masses, is  small
and therefore neglected.

As stated above, different hierarchies among the gaugino soft supersymmetry
breaking masses lead to different mixing scenarios for the neutralinos and the
charginos. However, this choice is constrained by  dark
matter  data. In order for LRSUSY models to feature a possible dark matter candidate,
the lightest supersymmetric particle has to be neutral. There are thus two
natural candidates, the lightest neutralino and the lightest (left-handed or
right-handed) sneutrino. In the MSSM, combining
cosmological and experimental collider constraints implies that phenomenologically viable scenarios with
(left-handed) sneutrino dark matter are difficult to achieve. On the one hand, a correct dark
matter relic density can  only be obtained  for very light or very heavy
sneutrinos, which prevents them from annihilating too fast into Standard Model
particles via a $Z$-boson-mediated diagram \cite{Ibanez:1983kw,Hagelin:1984wv,%
Falk:1994es}. On the other hand, very light sneutrinos are excluded by LEP data
on the invisible $Z$-boson width \cite{Beringer:2012zz} and very heavy
sneutrinos are excluded by experiments on dark matter direct detection
\cite{Falk:1994es}. In contrast, in LRSUSY, new possibilities open with a
possible right-handed sneutrino dark matter candidate which could account for
present data \cite{Arina:2008yh,Demir:2009kc}. However, this case  is not
considered in this work and we  require a neutralino to be the lightest
supersymmetric particle.

\begin{table}[!t]
\begin{center}
\begin{tabular}{l | r r r r}
  \hline\hline
  Parameter &  Scenario {\bf SI.1} & Scenario {\bf SI.2} & Scenario {\bf SII} & Scenario {\bf SIII}\\
  \hline
  \hline
  $M_1$ [GeV]    & 250  & 250  & 100 & 359\\
  $M_{2L}$ [GeV] & 500  & 750  & 250 & 320\\
  $M_{2R}$ [GeV] & 750  & 500  & 150 & 270\\
%
%
  \hline
  $v_R$ [GeV]       & 1000   & 1000   & 1300 & 1300\\
  $v_s$ [GeV]       & $10^5$ & $10^5$ & $10^5$ & $10^5$\\
  $\tan\beta$       & 10 & 10 & 10 & 10\\
  $\tan\tilde\beta$ & 1  & 1  & 1.05 & 1.05\\
\hline
  $\lambda_L$ & 0.1 & 0.1 & 0.1 & 0.1\\
  $\lambda_R$ & 0.1 & 0.1 & 0.1 & 0.1\\
  $\lambda_s$ & 0.1 & 0.1 & 0.1 & 0.1\\
  $\lambda_3$ & 0.1 & 0.1 & 0.1 & 0.1\\
 \hline\hline
\end{tabular}
\caption{\label{tab:bench} Benchmark scenarios allowing for different flavor
mixing and hierarchies among the neutralino and the chargino states. The slepton
masses $m_{\tilde L_L}$ and $m_{\tilde L_R}$ are kept free.}
\end{center}
\end{table}

We do not include any specific predictions for the Higgs masses. The Higgs sector of this variant of LRSUSY was studied in Ref.~\cite{Frank:2011jia}.
Although that analysis precedes the Higgs boson findings at the LHC and their parameter space differs somewhat from ours, some features are common for both. For $v_R$
of ${\cal O}$(TeV), the lightest CP-even neutral Higgs boson is basically the SM Higgs boson
and its mass and coupling parameters depend only on the bidoublet Higgs parameters. The mass is mostly affected by the coupling which generates $\tilde \mu_3$  ($\lambda_3$), and choosing $\lambda_3=0.1$ (as in our
benchmark scenarios) seems optimal for generating a SM Higgs mass around 125 GeV. Soft mass parameters in the Higgs scalar potential, absent from the chargino-neutralino mass matrices, ensure that the flavor-changing
neutral-current-mediating Higgs bosons are heavy. The Higgs mass analysis favors soft slepton masses
squared which are negative, while in our benchmark scenarios these are left as free parameters. Thus our choice of parameter space is consistent with a SM-like lightest neutral Higgs boson, for which a mass of 125 GeV can be obtained.

After inspecting the mass matrices of Eqs.\ \eqref{eq:mneu} and
\eqref{eq:mcharg} and recalling  that
we have chosen $v_s$ and $v_R$  very
large, only few options lead to a lightest supersymmetric particle
which is a neutralino. If the bino mass $M_1$ is smaller
than both wino masses, the lightest neutralino has a significant bino
component, which makes it subsequently lighter than the lightest chargino, the
latter being a wino state. However, in the case where $M_1$ is
larger than (at least) one of the two wino masses $M_{2L}$ and $M_{2R}$, the
mass difference between the lightest neutralino $\tilde\chi^0_1$ and chargino
$\tilde\chi^\pm_1$ is related to the bino fraction of the $\tilde\chi^0_1$
field. Therefore, $M_1$ has to be chosen small enough to guarantee enough
mixing, which consequently reduces the $\tilde\chi^0_1$ mass with
respect to the lightest chargino mass.

These considerations define our first two benchmark scenarios, denoted by
{\bf SI.1} and {\bf SI.2}, where we adopt three distinctly different gaugino masses,
the bino mass being the lightest. The full set of free parameters is
presented in the first two columns of Table \ref{tab:bench}.
Consequently, we deduce from the form of the matrices in Eqs.\ \eqref{eq:mneu} and
\eqref{eq:mcharg} that the mixing is drastically reduced, and that mass eigenstates
are almost purely gaugino-like.
This is illustrated in Figure \ref{fig:pure}, 
where we show the flavor decomposition of the five lighter neutralino and
chargino states, together with their mass.
The small value of the bino mass $M_1=250$ GeV  ensures that the lightest
supersymmetric particle is a bino state. We take wino masses of 500 GeV and 750
GeV, the $SU(2)_L$ wino mass being smaller in scenario {\bf SI.1}, and larger in the
second scenario {\bf SI.2}. This
hierarchy dictates  the flavor decomposition and masses 
of the four other chargino and neutralino states, as
depicted in the figure, where the $\W_L$ state is thus lighter (heavier) than
the $\W_R$ state in the upper (lower) panel of Figure~\ref{fig:pure}.

\begin{figure}
\begin{center}
	\includegraphics[width=.65\columnwidth]{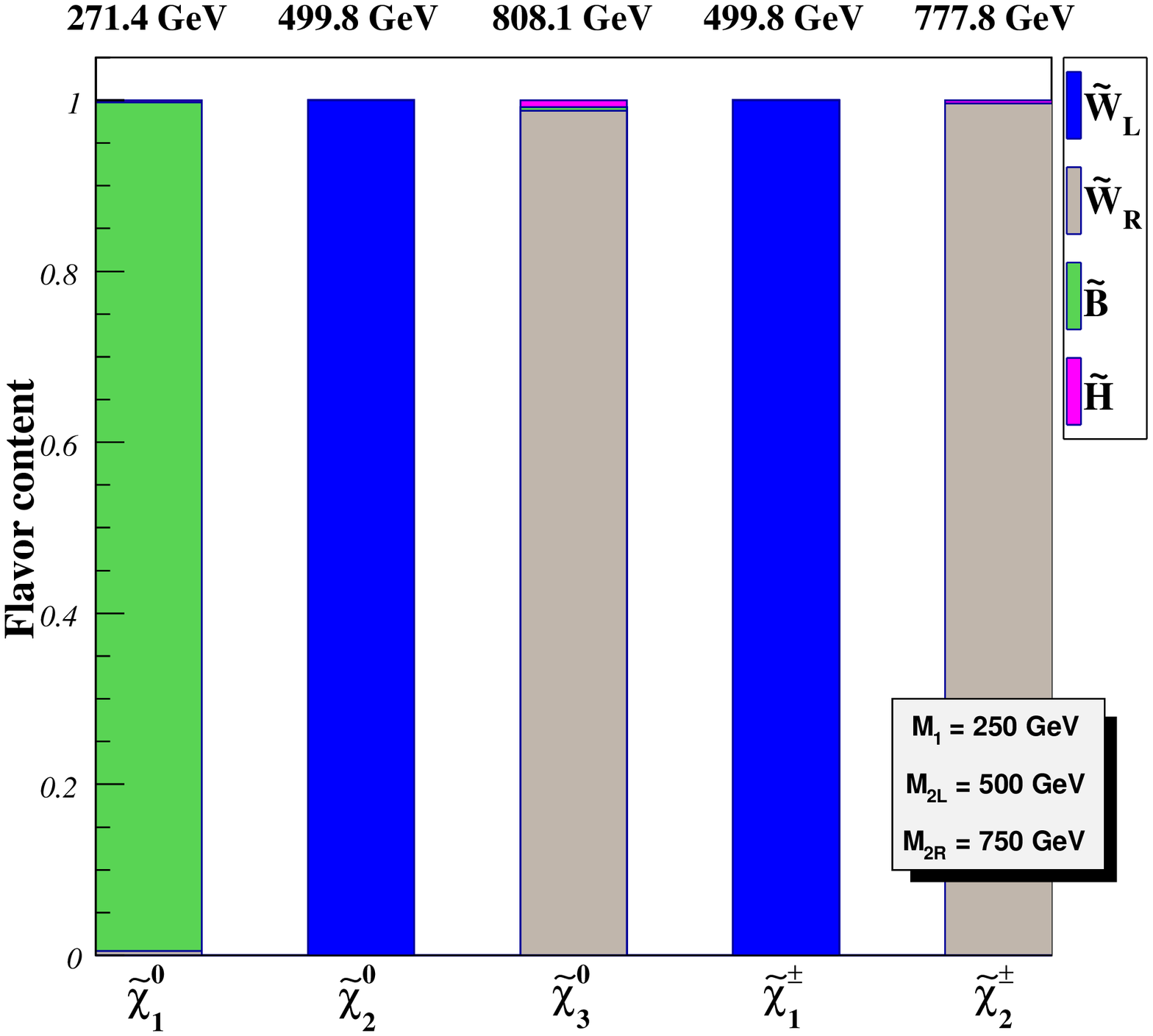} \\
	\includegraphics[width=.65\columnwidth]{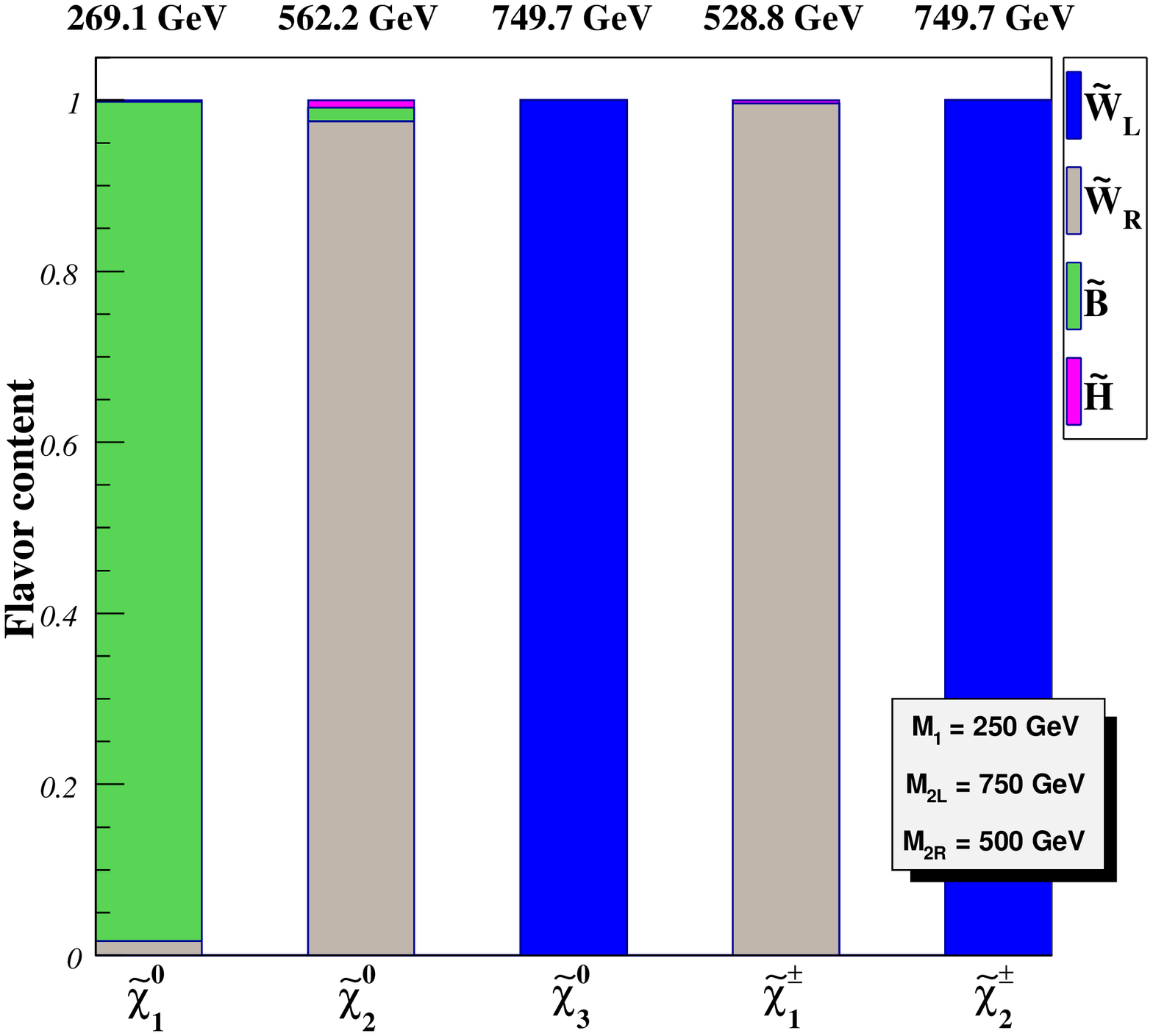}
\end{center}
\caption{Flavor decomposition and masses of the lighter neutralino and chargino
states for benchmark scenarios {\bf SI.1} (upper panel) and {\bf SI.2} (lower panel) as
defined in Table \ref{tab:bench}. The bino, $SU(2)_L$ and $SU(2)_R$ wino
components are presented in green, blue and gray, respectively, whilst the
higgsino component is shown in pink.}
\label{fig:pure}
\end{figure}

\begin{table}[t]
\begin{center}
\begin{tabular}{c||l|l|l}
\hline
\hline
Process & $~n=0~$ & $~n=1~$ & $~n=2~$\\
\hline 
\hline
$ \tchi_2^0   \to \tchi_1^0 + n~\ell+X\quad$  & 0.57 & 0.08 & 0.35 \\
\hline
$ \tchi_3^0   \to \tchi_1^0 + n~\ell+X\quad$ & 0.14 & 0.15 & 0.71 \\
\hline
$ \tchi_1^\pm \to \tchi_1^0 + n~\ell+X\quad$ & 0.22 & 0.78 & 0\\
\hline
$ \tchi_2^\pm \to \tchi_1^0 + n~\ell+X\quad$ & 0.22 & 0.78 & 0\\
\hline
\hline
\end{tabular}\\
\vspace{0.5cm}
\begin{tabular}{c||l|l|l}
\hline
\hline
Process & $~n=0~$ & $~n=1~$ & $~n=2~$\\
\hline 
\hline
$ \tchi_2^0   \to \tchi_1^0 + n~\ell+X\quad$ & 0.15 & 0.15 & 0.70 \\
$ \tchi_3^0   \to \tchi_1^0 + n~\ell+X\quad$ & 0.57 & 0.08 & 0.35 \\
$ \tchi_1^\pm \to \tchi_1^0 + n~\ell+X\quad$ & 0.22 & 0.78 & 0\\
$ \tchi_2^\pm \to \tchi_1^0 + n~\ell+X\quad$ & 0.22 & 0.78 & 0 \\
\hline
\hline
\end{tabular}
\caption{\label{tab:deca} Branching ratios of the lighter neutralinos and
charginos into charged leptons for scenario {\bf SI.1} (upper panel) and {\bf SI.2} (lower panel)
after fixing the slepton and sneutrino masses to
400~GeV. The decays of the intermediate tau
leptons, sleptons and sneutrinos are included. The symbol $X$ stands for
missing energy or jets.}
\end{center}
\end{table}

In these two non-mixing scenarios, 
winos decays are driven by the slepton masses $m_{\tilde L_L}$ and $m_{\tilde
L_R}$. As neutral and charged winos  originating from the same $SU(2)$ triplet
are almost mass-degenerate, a specific (neutral or charged) wino can 
only decay into sleptons
and sneutrinos of the corresponding chirality, together with the associated
Standard Model partner. Sleptons and sneutrinos further decay into the bino
state (the LSP), together with one additional lepton or neutrino, since their chirality
prevents them from decaying into the other wino state, even if kinematically allowed.
Depending on the difference between the wino and slepton
masses, the decay process consists either of a cascade of two two-body decays, 
or of a prompt three-body decay mediated by a virtual slepton or sneutrino.
This process
leads to at most two charged leptons produced in association with missing energy
related to the possible presence of
final state neutrinos and the one of the stable and invisible bino. This kind of cascade decay
is similar to those in the MSSM. The
relevant branching ratios of the lighter neutralinos and charginos to leptons
are indicated in Table~\ref{tab:deca}, assuming a universal slepton mass of 400~GeV.

For our third benchmark scenario, we choose a typical mixing scenario. In this case, $SU(2)_L$
winos are mostly pure, whilst 
$SU(2)_R$ winos mix significantly with the bino field. This mimics the neutral
and charged gauge boson mixing pattern presented in Appendix \ref{app:ewsb},
where $SU(2)_R \times U(1)_{B-L}$ is first broken to the hypercharge symmetry
group, implying a mixing of the $W_R^3$ and $\hat B$ gauge boson. In a similar fashion,
the  $\tilde W_R^3$ and $\tilde{\hat B}$ mix, which yields the hypercharge bino
$\tilde B'$. In a second step, the
electroweak gauge group is broken down to electromagnetism and the $\tilde B'$ field mixes with the $\tilde W_L^3$ field. As in the MSSM, this
mixing is in general rather small. This pattern is illustrated in scenario
{\bf SII}. The corresponding free parameters are presented in Table \ref{tab:bench}.
The Higgs sector parameters are fixed slightly differently from scenarios {\bf SI.1}
and {\bf SI.2} and we choose $M_1=100$ GeV, $M_{2L} = 250$ GeV and $M_{2R} = 150$
GeV,  the lower bino mass
ensuring a lightest
supersymmetric particle which is neutral. Although the resulting mass of the
lightest neutralino of 111~GeV and the one of the lightest chargino of about 200~GeV
seem ruled out by
current searches for electroweak superpartners at the LHC \cite{Aad:2012pxa,Aad:2012hba,ATLAS:2012kma,Aad:2012jja,
Chatrchyan:2012pka,Chatrchyan:2012mea,Chatrchyan:2011ff,CMS:aro},
the present constraints are evaded in the case of scenario {\bf SII}.
First, model independent searches always assume
a specific decay pattern (with a branching fraction of 100\%)
in the context of the MSSM. Next, already with a lightest neutralino
mass of more than 110~GeV, the searches lose sensitivity to lighter charginos
so that chargino masses of ${\cal O}(200~\text{GeV})$ are acceptable.

\begin{figure}
\begin{center}
	\includegraphics[width=.65\columnwidth]{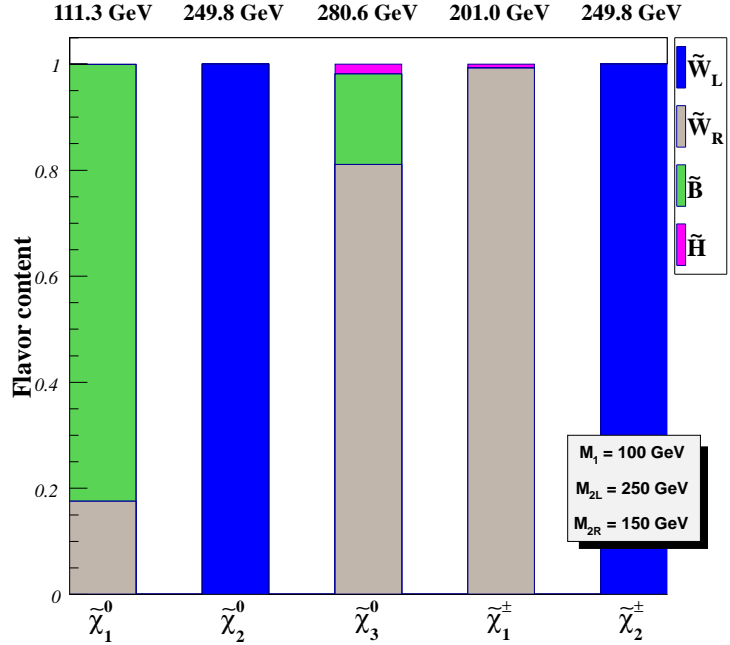} 
\end{center}
\caption{Same as in Figure \ref{fig:pure} but for our benchmark scenario {\bf SII}.}
\label{fig:mix1}
\end{figure}

The flavor decomposition of the five lighter
neutralino and chargino states is given in Figure
\ref{fig:mix1}, together with their mass eigenvalues. One observes a rather
important bino-right wino mixing (of order $20\%$) among the first and third
neutralino states. This opens new production channels for gaugino
pairs, since, \eg, the first chargino can now be produced in association with
both the first and the third neutralinos, and
new decay channels are also possible. For instance, 
the third neutralino can decay to the
lightest neutralino, in association with a $Z$-boson or a photon (the $Z^\prime$-boson
being too heavy), this new decay process 
 leading to a final state
with at most two charged leptons and a significant amount of 
missing energy. Furthermore, if the charged sleptons are lighter than the third
neutralino, the latter could also decay into a left-handed slepton-lepton pair. The produced
slepton decays further, producing another
lepton  and missing energy. The decay
patterns are illustrated in the upper and
lower panels of Table
\ref{tab:deca2}, where we show them for different universal slepton masses, chosen
equal to 200~GeV and 400~GeV, respectively\footnote{Sleptons with a 200~GeV mass are not excluded
by the most constraining LHC direct (and model-independent)
searches \cite{Aad:2012jja}. For a lightest neutralino with a mass of 111~GeV, the
region in the parameter space where the LHC starts to lose sensitivity corresponds exactly to the one where
sleptons are lighter than 200~GeV.}.
One observes than as soon as the charged current
decay channel of the third neutralino into a chargino is open, it becomes significant and reduces
the production of leptons from the SUSY particle decays.

\begin{table}[t]
\begin{center}
\begin{tabular}{c||l|l|l}
\hline
\hline
Process & $~n=0~$ & $~n=1~$ & $~n=2~$\\
\hline 
\hline
$ \tchi_2^0   \to \tchi_1^0 + n~\ell + X\quad$ & 0.57 & 0.08 & 0.35 \\
\hline
$ \tchi_3^0   \to \tchi_1^0 + n~\ell + X\quad$& 0.26 & 0.13 & 0.61 \\
\hline
$ \tchi_1^\pm \to \tchi_1^0 + n~\ell + X\quad $ & 1 & 0 & 0\\
\hline
$ \tchi_2^\pm \to \tchi_1^0 + n~\ell + X\quad $ & 0.22 & 0.78 & 0 \\
\hline
\hline
\end{tabular}\\
\vspace{0.5cm}
\begin{tabular}{c||l|l|l}
\hline
\hline
Process & $~n=0~$ & $~n=1~$ & $~n=2~$\\
\hline 
\hline
$ \tchi_2^0 \to \tchi_1^0 + n~\ell + X\quad $ & 0.57 & 0.08 & 0.35 \\
\hline
$ \tchi_3^0 \to \tchi_1^0 + n~\ell + X \quad $ & 0.12 & 0.09 & 0.43 \\
$ \tchi_3^0 \to \tchi_1^\pm  + n~\ell + X\quad $ & 0.35 & 0 & 0\\
\hline
$ \tchi_1^\pm \to \tchi_1^0 + n~\ell + X\quad  $ & 1 & 0 & 0\\
\hline
$ \tchi_2^\pm \to \tchi_1^0 + n~\ell + X\quad  $ & 0.22 & 0.78 & 0 \\
\hline
\hline
\end{tabular}
\caption{\label{tab:deca2} Branching ratios of the lighter neutralinos and
charginos into charged leptons for the scenario {\bf SII} with a universal slepton soft
mass of 200 (upper panel) and 400 (lower panel) GeV. The decays of the intermediate
tau lepton, sleptons and sneutrinos
are included. The symbol $X$ stands for missing energy or jets.}
\end{center}\end{table}

More interestingly, some specific hierarchies of the three soft gaugino
masses yield a large mixing in the neutralino sector. We design our
fourth and last benchmark point as a representative of these 
scenarios. With the choice of parameters presented in the last column of Table
\ref{tab:bench}, one obtains the flavor decomposition of the five lighter
neutralinos and charginos presented in Figure \ref{fig:mix2}.
In this case, the three soft gaugino masses are rather close to each other, \ie,
$M_1=359$ GeV, $M_{2L} = 320$ GeV and $M_{2R} = 270$ GeV, which leads to a significant mixing pattern. Moreover, two charginos and two neutralinos
are very close in mass, which could lead to displaced vertices,
the next-to-lightest neutralino and the two lighter charginos having a lifetime
of 4.13 ns, 0.09 ns and 1.09 ns, up to a possible boost factor, respectively. This
corresponds to decay lengths ranging from the order of the centimeter to the meter.
Furthermore, as in scenario {\bf SII}, such a gaugino hierarchy leads to
possibly lepton-enriched decay chains, although the produced leptons are expected
to be very soft due to the compression of the mass spectrum. The branching
ratios of the lighter neutralinos and charginos, for typical slepton masses of
400~GeV, are shown in Table \ref{tab:deca3}. Sleptons and
sneutrinos have hence been chosen heavier than most of the lighter neutralino and chargino states.
However, when real and not virtual, they decay further mainly to the lightest chargino (64\%) and to the second
lightest neutralino (36\%) whilst sneutrinos decay to the lightest chargino
(64\%) and to the lightest (21\%) and the next-to-lightest (15\%) neutralinos.

\begin{figure}
\begin{center}
	\includegraphics[width=.65\columnwidth]{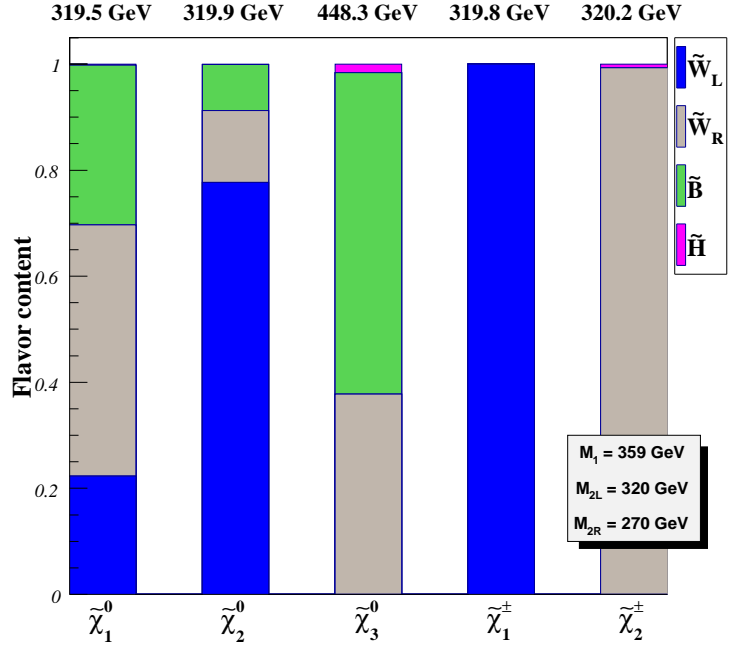}
\end{center}
\caption{Same as in Figure \ref{fig:pure} but for our benchmark scenario
{\bf SIII}.}
\label{fig:mix2}
\end{figure}

\begin{table}[t]
\begin{center}
\begin{tabular}{c||l|l|l}
\hline
\hline
Process & $~n=0~$ & $~n=1~$ & $~n=2~$\\
\hline 
\hline
$ \tchi_2^0 \to \tchi_1^0 + n~\ell   +X \quad $ & 0.22 & 0 & 0.16 \\
$ \tchi_2^0 \to \tchi_1^\pm+ n~\ell  +X \quad $ & 0.52 & 0.10 &0 \\
\hline 
$ \tchi_3^0 \to \tchi_1^0 + n~\ell   +X \quad $ & 0.10 & 0 & 0 \\
$ \tchi_3^0 \to \tchi_2^0 + n~\ell   +X \quad $ & 0.10 & 0.03 &  0.13 \\
$ \tchi_3^0 \to \tchi_1^\pm  + n~\ell+X \quad $ & 0.14 & 0.51 &  0\\
\hline 
$ \tchi_1^\pm \to \tchi_1^0 + n~\ell +X \quad $ & 0.84 & 0.16&0  \\
\hline 
$ \tchi_2^\pm \to \tchi_1^0 + n~\ell +X \quad $  & 0.996 & 0 & 0 \\
$ \tchi_2^\pm \to \tchi_2^0 + n~\ell +X \quad $  & 0.004 & 0 & 0 \\
\hline 
\hline 
\end{tabular}
\caption{\label{tab:deca3} Branching ratios of the lighter  neutralinos and
charginos into charged leptons for the scenario {\bf SIII} with a universal slepton soft
mass of 400 GeV.  The decays of the intermediate tau
leptons, sleptons and sneutrinos are included. The
symbol $X$ stands for missing energy or jets.}
\end{center}
\end{table}


\section{Gauginos as probes of left-right symmetric supersymmetry
at the LHC} \label {sec:production}

\subsection{General considerations}
\begin{table}[t]
\begin{center}
\begin{tabular}{c || c | c | c}
\hline
\hline
Process & $~$ $\sqrt{S} = 7$ TeV [fb] $~$ & $~$ $\sqrt{S} = 8$ TeV [fb]$~$ & $~$ $\sqrt{S} = 14$ TeV [fb]$~$\\ 
\hline 
\hline
$p p \to \tchi_2^0 \tchi_1^\pm\quad$  & $\quad 13.2\quad$ & $\quad 20.6\quad$
  &$\quad  89.6\quad$ \\
$p p \to \tchi_3^0 \tchi_2^\pm\quad$  & $\quad 0.71\quad$ & $\quad 1.40\quad$
  &$\quad  11.4\quad$ \\
$p p \to \tchi_1^0 \tchi_2^\pm\quad$  & $\quad < 0.1\quad$ & $\quad< 0.1\quad$
  &$\quad  0.39\quad$ \\
\hline
$p p \to \tchi_1^+ \tchi_1^-\quad$  & $\quad 2.90\quad$ & $\quad 4.61\quad $ &
  $\quad 21.2\quad $ \\
$p p \to \tchi_2^+ \tchi_2^-\quad$  & $\quad 0.21\quad$ & $\quad 0.42\quad $ &
  $\quad 3.41 \quad$\\
\hline
\hline
\end{tabular} \\
\vspace{0.5cm}
\begin{tabular}{c || c | c | c}
\hline
\hline
Process & $~$ $\sqrt{S} = 7$ TeV [fb] $~$ & $~$ $\sqrt{S} = 8$ TeV [fb]$~$ & $~$ $\sqrt{S} = 14$ TeV [fb]$~$\\ 
\hline 
\hline
$p p \to \tchi_2^0 \tchi_1^\pm\quad$  & $\quad 10.2 \quad$ & $\quad 16.3 \quad$
  &$\quad 76.0 \quad$ \\
$p p \to \tchi_3^0 \tchi_2^\pm\quad$  & $\quad 0.98 \quad$ & $\quad 1.86\quad$
  &$\quad 13.8 \quad$ \\
$p p \to \tchi_1^0 \tchi_1^\pm\quad$  & $\quad 1.16 \quad$ & $\quad 1.67\quad$
  &$\quad 5.88 \quad$ \\
\hline
$p p \to \tchi_1^+ \tchi_1^-\quad$  & $\quad 4.49 \quad$ & $\quad 7.13 \quad $ &
  $\quad 32.9 \quad $ \\
$p p \to \tchi_2^+ \tchi_2^-\quad$  & $\quad 0.21 \quad$ & $\quad 0.40 \quad $ &
  $\quad 3.14 \quad$\\
\hline
\hline
\end{tabular}
\caption{\label{tab:prod1} Dominant cross sections, given in fb and at
the leading order of perturbative QCD, of neutralino and
chargino pair production at the LHC for a center-of-mass energy of 7
TeV, 8 TeV and 14 TeV. Results are shown for  benchmark scenarios {\bf SI.1} (upper panel)
and {\bf SI.2} (lower panel) after setting the slepton masses to a universal value of 400 GeV.}
\end{center}
\end{table}

\begin{table}[t]
\begin{center}
\begin{tabular}{c || c | c | c}
\hline
\hline
Process & $~$ $\sqrt{S} = 7$ TeV [fb] $~$ & $~$ $\sqrt{S} = 8$ TeV [fb]$~$ & $~$ $\sqrt{S} = 14$ TeV [fb]$~$\\ 
\hline 
\hline
$p p \to \tchi_3^0 \tchi_1^\pm\quad$  & $\quad 4999 \quad$ & $\quad 6530 \quad$
  &$\quad 17490 \quad$ \\
$p p \to \tchi_1^0 \tchi_1^\pm\quad$  & $\quad 3139 \quad$ & $\quad 4085\quad$
  &$\quad 10830 \quad$ \\
$p p \to \tchi_2^0 \tchi_2^\pm\quad$  & $\quad 387 \quad$ & $\quad 514 \quad$
  &$\quad 1452 \quad$ \\
$p p \to \tchi_2^0 \tchi_1^\pm\quad$  & $\quad 0.83 \quad$ & $\quad 1.09\quad$
  &$\quad 2.88\quad$ \\
$p p \to \tchi_3^0 \tchi_2^\pm\quad$  & $\quad < 0.1 \quad$ & $\quad < 0.1\quad$
  &$\quad 0.11 \quad$ \\
\hline
$p p \to \tchi_1^+ \tchi_1^-\quad$  & $\quad 532 \quad$ & $\quad 780\quad $ &
  $\quad 2851\quad $ \\
$p p \to \tchi_2^+ \tchi_2^-\quad$  & $\quad 92.2 \quad$ & $\quad 123 \quad $ &
  $\quad 355.9 \quad$\\
\hline
\end{tabular}\\
\vspace{0.5cm}
\begin{tabular}{c || c | c | c}
\hline
\hline
Process & $~$ $\sqrt{S} = 7$ TeV [fb] $~$ & $~$ $\sqrt{S} = 8$ TeV [fb]$~$ & $~$ $\sqrt{S} = 14$ TeV [fb]$~$\\ 
\hline 
$p p \to \tchi_3^0 \tchi_1^\pm\quad$  & $\quad 5188 \quad$ & $\quad 6776 \quad$
  &$\quad 18140 \quad$ \\
$p p \to \tchi_1^0 \tchi_1^\pm\quad$  & $\quad 3255 \quad$ & $\quad 4236\quad$
  &$\quad 11230 \quad$ \\
$p p \to \tchi_2^0 \tchi_2^\pm\quad$  & $\quad 387 \quad$ & $\quad 514 \quad$
  &$\quad 1451 \quad$ \\
$p p \to \tchi_2^0 \tchi_1^\pm\quad$  & $\quad 0.86 \quad$ & $\quad 1.13\quad$
  &$\quad 3 \quad$ \\
$p p \to \tchi_3^0 \tchi_2^\pm\quad$  & $\quad < 0.1 \quad$ & $\quad < 0.1\quad$
  &$\quad 0.11 \quad$ \\
\hline
$p p \to \tchi_1^+ \tchi_1^-\quad$  & $\quad 572 \quad$ & $\quad 838\quad $ &
  $\quad 3059\quad $ \\
$p p \to \tchi_2^+ \tchi_2^-\quad$  & $\quad 92.2 \quad$ & $\quad 123 \quad $ &
  $\quad 356 \quad$\\
\hline
\end{tabular}\\
\vspace{0.5cm}
\begin{tabular}{c || c | c | c}
\hline
\hline
Process & $~$ $\sqrt{S} = 7$ TeV [fb] $~$ & $~$ $\sqrt{S} = 8$ TeV [fb]$~$ & $~$ $\sqrt{S} = 14$ TeV [fb]$~$\\ 
\hline 
\hline
$p p \to \tchi_2^0 \tchi_1^\pm\quad$  & $\quad 99.6 \quad$ & $\quad 137\quad$
  &$\quad 433 \quad$ \\
$p p \to \tchi_1^0 \tchi_2^\pm\quad$  & $\quad 93.6 \quad$ & $\quad 128\quad$
  &$\quad 393 \quad$ \\
$p p \to \tchi_1^0 \tchi_1^\pm\quad$  & $\quad 28.5 \quad$ & $\quad 39.3 \quad$
  &$\quad 125 \quad$ \\
$p p \to \tchi_2^0 \tchi_2^\pm\quad$  & $\quad 26.7 \quad$ & $\quad 36.6 \quad$
  &$\quad 113 \quad$ \\
$p p \to \tchi_3^0 \tchi_2^\pm\quad$  & $\quad 13.0 \quad$ & $\quad 19.0\quad$
  &$\quad 69.2 \quad$ \\
\hline
$p p \to \tchi_2^+ \tchi_2^-\quad$  & $\quad 537 \quad$ & $\quad 788 \quad $ &
  $\quad 2887\quad$\\
$p p \to \tchi_1^+ \tchi_1^-\quad$  & $\quad 29.8 \quad$ & $\quad 41.7\quad $ &
  $\quad 137 \quad $ \\
\hline
\hline
\end{tabular}
	\caption{\label{tab:prod2} Same as in Table \ref{tab:prod1}, but for 
benchmark scenarios {\bf SII} after setting the slepton masses to a universal value of
200 GeV (upper panel) and 400 GeV (middle panel) and for  scenario {\bf SIII} with a 
universal slepton mass of 400 GeV (lower panel).}
\end{center}
\end{table}

At the LHC, neutralinos $\tilde{\chi}^{0}$ and charginos
$\tilde{\chi}^\pm$ can be produced directly in pairs or in association with
gluinos $\tilde{g}$ or with squarks $\tilde{q}$. In the scenarios
considered in this work, squarks and gluinos are very heavy and decoupled.
Therefore, the only relevant production processes are
\be \label{ino-production}
    p\ p \rightarrow 
     \tilde{\chi}_i^\mp \tilde{\chi}_j^\pm\ , \quad
     \tilde{\chi}_i^0 \tilde{\chi}_j^0\quad\text{and}\quad 
     \tilde{\chi}_i^0 \tilde{\chi}_j^\pm \ ,
\ee
via $s$-channel gauge boson exchange or $t/u$-channel squark exchange.
 Although existing bounds on
the $Z_R$ and $W_R$ boson masses \cite{Beringer:2012zz} force them to be heavy, pair production
of the neutralino and chargino states through one of the new vector
boson is not necessarily suppressed at the LHC energies, as the cross section
may get enhanced by important  (although not considered in this work)
resonance effects. In addition, the decoupling  of 
squarks also leads to an increase in the cross section, taming the
destructive interferences between the $s$- and $t/u$-channel diagrams.
In Table \ref{tab:prod1} and Table \ref{tab:prod2}  we present 
numerical predictions for the most relevant associated production cross
sections at the LHC, running at a center-of-mass energy of 7 TeV (2010-2011 run), 8 TeV (2012
run) and 14 TeV (future run). Numerical computations have been performed using 
the matrix element generator {\sc MadGraph} 5
\cite{Alwall:2011uj}, after convoluting the produced hard-scattering squared matrix elements
with the leading order set of parton densities CTEQ6L1 \cite{Pumplin:2002vw}.
The LRSUSY UFO files \cite{Degrande:2011ua} necessary for {\sc MadGraph}~5
have been generated
with the program {\sc FeynRules} \cite{Christensen:2008py, Christensen:2009jx,
Christensen:2010wz, Duhr:2011se, Fuks:2012im, Alloul:2013fw} after  implementing the
Lagrangian introduced in Section \ref{sec:model}. The results shown in
the tables correspond to a factorization scale fixed to the transverse mass of the produced
superparticles and are given at the leading-order of perturbative QCD. We omit all channels
involving a cross section smaller than 0.1~fb and set the slepton and sneutrino masses  to 400~GeV,
unless otherwise stated.

Considering neutralino and chargino production in the MSSM, more precise calculations are available
once we add next-to-leading order corrections
\cite{Beenakker:1999xh,Debove:2010kf} combined with the resummation of the leading and next-to-leading logarithms
to all orders in the strong coupling \cite{Debove:2009ia, Debove:2010kf, Debove:2011xj}. This is known to increase the
cross sections by about $20\%-25\%$. Although no such precision computations have been achieved in the LRSUSY
context, the structure of the next-to-leading order calculations is very similar in both the MSSM and the LRSUSY cases.
We therefore adopt, in the rest of this paper, next-to-leading values to be equal to leading-order results for LRSUSY signal cross sections multiplied
by constant $K$-factor fixed to $1.20$.

Once produced, all neutralinos and charginos decay into isolated leptons,
hard jets and missing energy carried by the LSPs by means of cascades of
two-body (and possibly three-body)
decays. We choose to focus on the pattern with
the cleanest collider signature, when several hard isolated leptons are produced.
For instance, a typical gaugino cascade decay would be
\be\label{decay}\bsp
   \mbox{(heavy chargino/neutralino)} & \rightarrow \mbox{(lepton)}\;
     \mbox{(slepton)}^{(\star)} \\ & \rightarrow \mbox{(lepton)}\; \mbox{(lepton)}\;
     \mbox{(light chargino/neutralino)}\ , 
\esp\ee
where the (slepton)$^{(\star)}$ is a real (virtual) slepton, as lighter (heavier) than the
heavy chargino or neutralino. Channels with intermediate gauge bosons are sometimes
open and lead to similar final state signatures.
Although such cascades exist in both MSSM and LRSUSY models,
an explicit analysis of the neutralino and chargino
production and decays could be useful to unveil differences between the two symmetries.
In principle, while the final state signatures are very similar in terms of
number of leptons, jets and missing energy, the intermediate decay stages could be
different, yielding different results in terms of branching ratios
and kinematical distributions.

\renewcommand{\arraystretch}{1.4}
\begin{table}[t]
\begin{tabular}{l l l}
    \hline\hline 
   Process $\quad$ &Signature$\qquad\qquad$ &  Representative candidate processes \\
    \hline\hline
    I. & $0 \, \ell + \slashed{E}_T$ & 
     $p\, p \rightarrow \tilde{\chi}_1^0  \tilde{\chi}_1^0$ \\ 
\hline 
    II. & $1\, \ell + \slashed{E}_T$ & 
      $p\, p \rightarrow \left(\tilde{\chi}_1^\pm \rightarrow
      \ell^\pm{\nu}_{\ell} \tilde{\chi}_1^0\right)\, \tilde{\chi}_1^0$ \\ 
\hline
  III.A & \multirow{2}{*}{$2\, \ell + \slashed{E}_T$} & 
   $ p\, p \rightarrow \left(\tilde{\chi}_2^0 \rightarrow \ell^\pm {\ell}^\mp
   \tilde{\chi}_1^0\right)\, \tilde{\chi}_1^0 $ \\ 
  III.B & & $ p\, p \rightarrow
    \left(\tilde{\chi}_1^\pm \rightarrow \ell^\pm {\nu}_{\ell}
    \tilde{\chi}_1^0\right)\, \left(\tilde{\chi}_1^{\mp} \rightarrow
    \ell^{\prime \mp} {\nu}_{\ell^\prime} \tilde{\chi}_1^0\right)$\\ 
\hline
  IV.& $ 3\, \ell + \slashed{E}_T$ & $ p\, p \rightarrow
    \left(\tilde{\chi}_2^0 \rightarrow \ell^\pm {\ell}^\mp
    \tilde{\chi}_1^0\right)\,\left(\tilde{\chi}_1^\pm \rightarrow
    \ell^{\prime \pm} {\nu}_{\ell^\prime} \tilde{\chi}_1^0\right)$\\ 
\hline 
  V.A & \multirow{2}{*}{$ 4\, \ell + \slashed{E}_T$} & $p\, p \rightarrow
    \left(\tilde{\chi}_2^0 \rightarrow \ell^\pm {\ell}^\mp
    \tilde{\chi}_1^0\right)\, \left(\tilde{\chi}_2^0 \rightarrow
    \ell^{\prime \pm} {\ell}^{\prime \mp} \tilde{\chi}_1^0\right)\,$\\
  V.B & & $p\, p \rightarrow
       \left(\tilde{\chi}_1^{\mp} \rightarrow
       \ell^\mp {\nu}_\ell \tilde{\chi}_1^0\right)\, 
       \left(\tilde{\chi}_3^0 \rightarrow 
       \ell^{\prime \pm} \ell^{\prime\prime \pm}
       {\ell}^{\prime\prime \mp} {\nu}_{\ell^\prime} j j \tilde{\chi}_1^0\right)$\\
\hline 
  VI.& $ 5\, \ell + \slashed{E}_T$ & $p\, p \rightarrow
    \left(\tilde{\chi}_3^0 \rightarrow \ell^\pm {\ell}^\mp 
      \ell^{\prime \pm} {\ell}^{\prime \mp} \tilde{\chi}_1^0\right)\, 
      \left(\tilde{\chi}_1^\pm \rightarrow \ell^{\prime\prime \pm}
      \nu_{\ell^{\prime\prime}} \tilde{\chi}_1^0\right)$\\
\hline 
  VII. & $6\, \ell + \slashed{E}_T$ & $p\, p \rightarrow
    \left(\tilde{\chi}_3^0 \rightarrow \ell^\pm {\ell}^\mp 
      \ell^{\prime \pm} {\ell}^{\prime \mp} \tilde{\chi}_1^0\right)\, 
      \left(\tilde{\chi}_2^0 \rightarrow \ell^{\prime\prime \pm}
     {\ell}^{\prime\prime
      \mp} \tilde{\chi}_1^0\right)\,$\\
\hline 
  VIII. & $ 7\, \ell + \slashed{E}_T$ & $p\, p \rightarrow
    \left(\tilde{\chi}_3^0 \rightarrow \ell^\pm {\ell}^\mp 
      \ell^{\prime \pm} {\ell}^{\prime \mp} \tilde{\chi}_1^0\right)\,
       \left(\tilde{\chi}_3^0 \rightarrow 
       \ell^{\prime\prime \pm} \ell^{\prime\prime\prime \pm}
       {\ell}^{\prime\prime\prime \mp} {\nu}_{\ell^{\prime\prime}} j j \tilde{\chi}_1^0\right)
       $ \\
\hline 
   IX. & $ 8\, \ell + \slashed{E}_T$ & $p\, p \rightarrow
    \left(\tilde{\chi}_3^0 \rightarrow \ell^\pm {\ell}^\mp 
      \ell^{\prime \pm} {\ell}^{\prime \mp} \tilde{\chi}_1^0\right)\,
      \left(\tilde{\chi}_3^0 \rightarrow \ell^{\prime\prime\pm}
      {\ell}^{\prime\prime\mp} 
      \ell^{\prime\prime\prime \pm} {\ell}^{\prime\prime\prime \mp}
      \tilde{\chi}_1^0\right)$\\
\hline\hline
\end{tabular}
\caption{\label{table3} Multilepton LHC signatures related to chargino and
  neutralino production and decays in LRSUSY models.
 As each type of signature receives contributions
  from one or more decay processes, we only give a few representative examples.}
\end{table}
\renewcommand{\arraystretch}{1.}

In the MSSM, gaugino cascade decays as above involve both hypercharge
and/or
$SU(2)_L$ gauginos\footnote{We recall that we only consider cases where
the higgsino fields are decoupled.}. In
LRSUSY models, the field content is supplemented by new neutral and charged gauge
fermions, and the hypercharge bino state originates from the mixing of the
$B-L$ bino and $SU(2)_R$ neutral wino.
Gaugino decay chains could hence acquire novel features not present in the MSSM. For
example, we start from the MSSM decay
\be \label{w3decay-MSSM}
  \tilde{W}_L^{3} \rightarrow \ell_L^+ \tilde{\ell}_L^{\star -}
    \rightarrow \ell^{+} \ell^{-} \tilde{B}'\ .
\ee
As the $\tilde{W}_L^3$ and $\tilde{W}_L^{\pm}$ fields are nearly
mass-degenerate since $SU(2)_L$ breaking splitting effects are
small, gaugino-to-gaugino decays are hardly
possible. The winos then mostly decay through (virtual or real) sleptons, which  yields
signatures with at most two charged leptons.
In contrast, LRSUSY gaugino-to-gaugino decays
are possible, as for example in the case of the benchmark scenarios {\bf SII} and {\bf SIII}
(see Table
\ref{tab:deca2} and Table~\ref{tab:deca3}), where the physical
states are admixtures of different gaugino states (see
Figure~\ref{fig:mix1} and Figure~\ref{fig:mix2}).
This leads to lepton-enriched decay chains such as
\be \label{w3decay-U1p}
  \tilde{W}^0_R \rightarrow 
   \ell_R^+ \tilde{\ell}_R^{(\star) -} \rightarrow 
   \ell_R^+ \ell_R^- \tilde{\chi}_2^0 \rightarrow 
   \ell_R^+ \ell_R^- {\ell}_L^{\prime\, +} \tilde{\ell}_L^{(\star) \prime\,-} \rightarrow
   \ell_R^+ \ell_R^- {\ell}_L^{\prime\, +} {\ell}_L^{\prime\, -}
  \tilde{\chi}^0_1 \ ,
\ee
where the two lighter neutralinos have both $B-L$ bino and $SU(2)_L$ wino components. In this case, the decay
yields a tetralepton plus missing energy final state.
Whereas allowing for immediately distinguishing LRSUSY models from the MSSM, these types of signatures
suffer from very low branching fractions in our scenarios and,   depending on
final state lepton hardness, could be difficult to observe.

We generically classify the different final state signatures possibly arising
from the production and decay of two gauginos in LRSUSY models in Table \ref{table3}.
Our classification is based on the number of produced leptons determined by the production
and decay of the five LRSUSY gaugino states.
For each type of signature, we give one
representative associated process. In contrast to the MSSM
where the number of produced leptons is limited to at most four
(considering only the lightest chargino and the two lighter neutralinos),
LRSUSY models feature the production of final states with up to eight charged
leptons, a Standard Model background free signature if the associated production rate
is large enough.

In the rest of this section, we first (Section~\ref{sec:lrsusy}) focus on the possible extraction
of a LRSUSY signal from the SM background in the context
of the benchmark scenarios of Section~\ref{sec:benchmarks}.
Next, in the eventuality of the observation of an excess of leptonic events at the LHC, we show, for similar mass spectra in the MSSM and LRSUSY cases, several ways to disentangle both models
(Section~\ref{sec:mssm}).

\subsection{Leptonic signatures of left-right symmetric gauginos at the LHC}
\label{sec:lrsusy}

We now concentrate on phenomenological analyses relying on Monte Carlo simulations of collisions produced at
the LHC, for a center-of-mass energy of 8~TeV and an integrated luminosity of 20~fb$^{-1}$.
For both signal and background, we make use of the
{\sc MadGraph}~5~\cite{Alwall:2011uj} package for generating hard process matrix elements,
including up to two additional jets for Standard Model contributions,  convoluted with the parton density set
CTEQ6L1~\cite{Pumplin:2002vw}. Parton-level events were then
 matched to parton showering and hadronization by means of the program
{\sc Pythia}~\cite{Sjostrand:2006za,Sjostrand:2007gs} and merged
according to the $k_T$-MLM scheme \cite{Mangano:2006rw,Alwall:2008qv}.
Focusing on leptonic final states, we assumed perfect electron and muon reconstruction.
This is a fair approximation when one accounts for appropriate object selection criteria
based on, \eg, large transverse momentum. Simulated events were eventually  analyzed within
the {\sc MadAnalysis}~5~framework~\cite{Conte:2012fm}.

We  generated dedicated event samples for various sources of Standard Model background and
reweighted the samples according to calculations for the total  production rates
convoluting next-to-leading order (NLO) or even
next-to-next-to-leading order (NNLO) matrix elements, when available,
with the CT10 parton densities~\cite{Lai:2010vv}.
Events originating from the leptonic or invisible decay of a $W$-boson or $Z$-boson produced
in association with jets have been reweighted to the NNLO accuracy, using total rates of 35678~pb and 10319~pb,
respectively, as predicted by the {\sc Fewz} program~\cite{Gavin:2012sy,Gavin:2010az}.
Inclusive top-antitop events have been normalized to
a cross section of 255.8~pb, as derived by the {\sc Hathor} package~\cite{Aliev:2010zk}, which includes all
NLO diagrams and genuine NNLO contributions, while
single top event generation in the $t$-, $tW$- and $s$-channel topology has been
normalized, at an approximate NNLO accuracy,
to 87.2~pb, 22.2~pb and 5.5~pb, respectively~\cite{Kidonakis:2012db}.
Diboson events have been rescaled to a weight derived from
NLO results as computed by means of the {\sc Mcfm} software, using cross sections of
30.2~pb, 11.8~pb and 4.5~pb for the $WW$, $WZ$ and $ZZ$ channels,
respectively~\cite{Campbell:1999ah,Campbell:2011bn,Campbell:2012dh}.
In this case, fully hadronic decay modes have been neglected.
Next, $ttW$ and $ttZ$ events were normalized to
NLO, using again {\sc Mcfm}, while the normalization of the other simulated rare Standard Model
processes relied on {\sc MadGraph}~5 results. We hence employed cross sections of
0.25~pb, 0.21~pb, 46~fb, 13~fb and 0.7~fb, for the
$ttW$, $ttZ$, $tZj$, $ttWW$ and $tttt$ channels, respectively.
Finally, we did not consider
multijet events, their correct treatment requiring data-driven methods. However,
basing the analyses of this work on final states containing charged leptons with very large transverse momentum
and a sensible quantity of missing transverse energy,
we expect those contributions to be fully under control \cite{Aad:2012wm,Chatrchyan:2012oaa}.

We designed three analyses possibly sensitive to LRSUSY signals based on different event lepton multiplicity
and subsequently divided the background and event
samples in three categories. We separately considered events
with one single lepton (see Section~\ref{sec:monol}), two leptons (see Section~\ref{sec:dil})
and more than two leptons (see Section~\ref{sec:multil}). This distinction was made after
defining jet and lepton candidates as follows.
\begin{itemize}
  \item Jets are reconstructed by means of
    the {\sc FastJet} program~\cite{Cacciari:2005hq,Cacciari:2011ma}, using an anti-$k_{t}$
    algorithm of radius parameter $R=0.4$ \cite{Cacciari:2008gp}.
  \item We only retain jet candidates if their transverse momentum is greater than 20~GeV
    and their pseudorapidity fulfills $|\eta| \leq 2.5$.
  \item We select electrons and muons candidates having
    a transverse momentum $p_T$ larger than 10~GeV and a pseudorapidity $|\eta| \leq 2.5$.
  \item We remove jet objects which lie in an angular distance  $\Delta R  = \sqrt{\Delta\phi^2 + \Delta\eta^2}
    \leq 0.1$ of an electron, where $\phi$ stands for the azimuthal angle with respect to the beam
    direction.
  \item We remove electrons and muons lying in a cone of radius $\Delta R \leq 0.4$ of any of the
    remaining jets.
\end{itemize}
We then vetoed events containing at least one $b$-tagged jet, including a $b$-tagging efficiency of 60\%
for a charm/light mistagging rate of 10\%/1\%.

\subsubsection{A single lepton signature}\label{sec:monol}
As deduced from the branching ratio tables computed
in Section~\ref{sec:benchmarks}, LRSUSY gaugino pair production at the LHC are likely to give rise to
events containing exactly one charged lepton and
a sensible amount of missing transverse energy carried by the undetected LSPs.
The associated production cross section is however rather reduced in many cases, as shown in Table~\ref{tab:prod1}
and Table~\ref{tab:prod2} for the benchmark scenarios under consideration,
which may render the observation of any LRSUSY hint challenging. In this analysis, we select events with exactly $N_\ell = 1$
charged lepton. After applying the object definitions above-mentioned, the signal efficiency ranges from less than 1\% in the
case of the {\bf SIII} scenario, where most of the leptons are too soft to be detected due to small
splittings in the mass spectrum (see Figure~\ref{fig:mix2}), to 42\% for the scenario {\bf SI.1}.
At this stage, the SM background overwhelms the signal by more than four orders of magnitude and
is dominated by $W$+jets events (94\%) and $Z$+jets events (5.7\%), where
one of the lepton either lies outside the $\eta\leq 2.5$ region, or is too soft for being observed (with $p_T<10$~GeV),
or is non-isolated. We recall that those numbers do not include
non-simulated multijet background events possibly yielding final state signatures with charged leptons originating
from the hadronization process.
\begin{figure}
\begin{center}
   \includegraphics[width=.78\columnwidth]{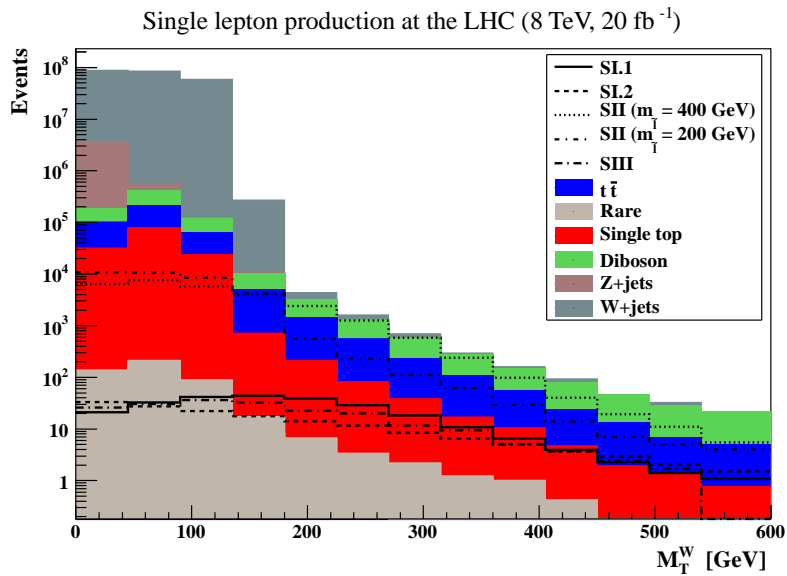}
 \end{center}
 \caption{\label{fig:monolep_MT} Distribution of the $M_T^W$ variable defined in Eq.~\eqref{eq:mtw} after
  selecting events with exactly one charged lepton and vetoing events with at least one
  $b$-tagged jet. We considered 20 fb$^{-1}$ of LHC collisions at a center-of-mass energy of 8~TeV and present results for the
  different background contributions and for all the considered LRSUSY scenarios.}
\end{figure}

\begin{figure}
\begin{center}
  \includegraphics[width=.78\columnwidth]{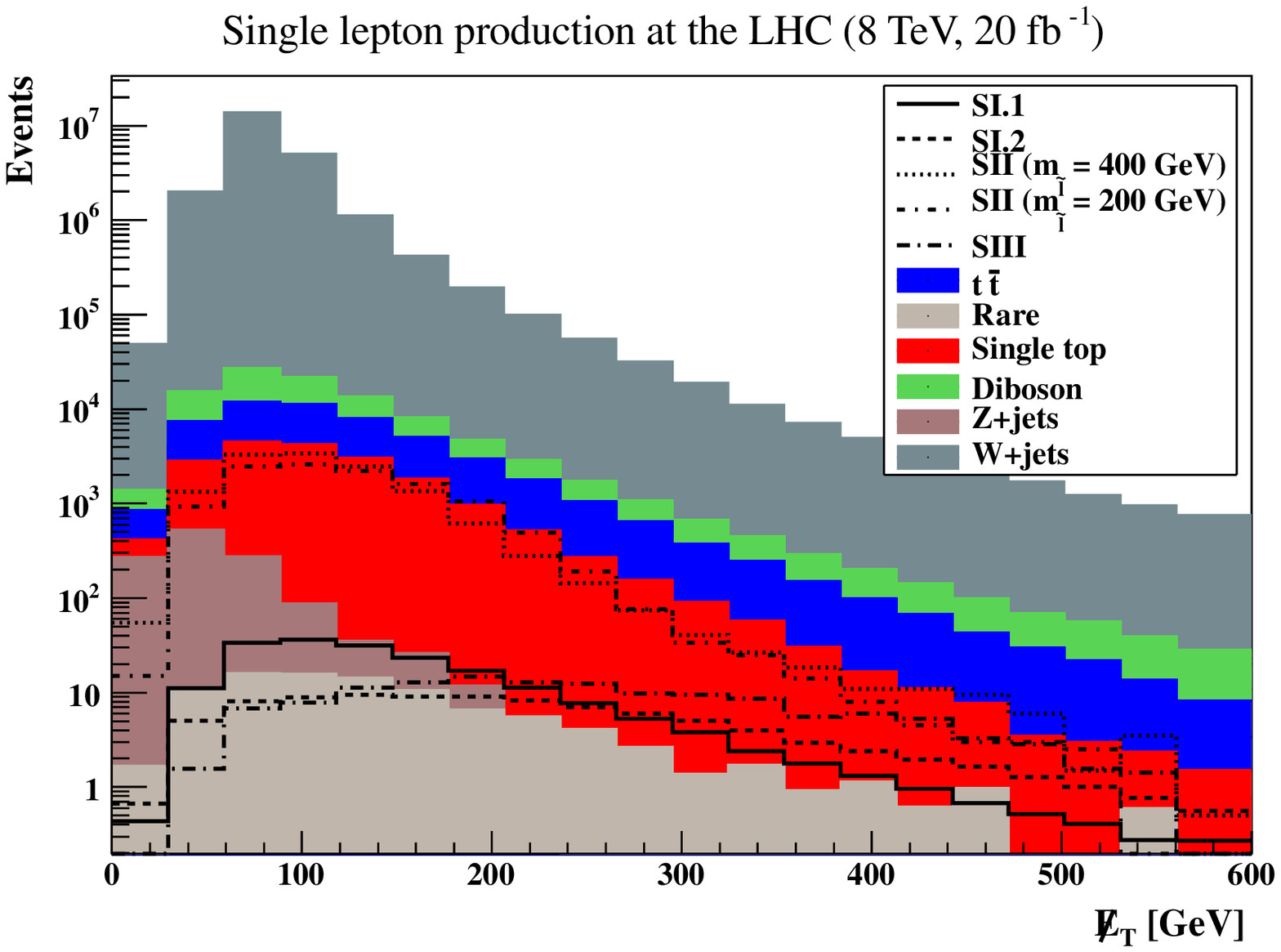}
 \end{center}
 \caption{\label{fig:monolep_MET} Missing transverse energy distribution after vetoing events with at least one
  $b$-tagged jet and selecting events with exactly one charged lepton and $M_T^W \geq 100$~GeV.
  We considered 20 fb$^{-1}$ of LHC collisions at a center-of-mass energy of 8~TeV and present results for the
  different background contributions and for all the considered LRSUSY scenarios.}
\end{figure}

\begin{figure}
\begin{center}
  \includegraphics[width=.78\columnwidth]{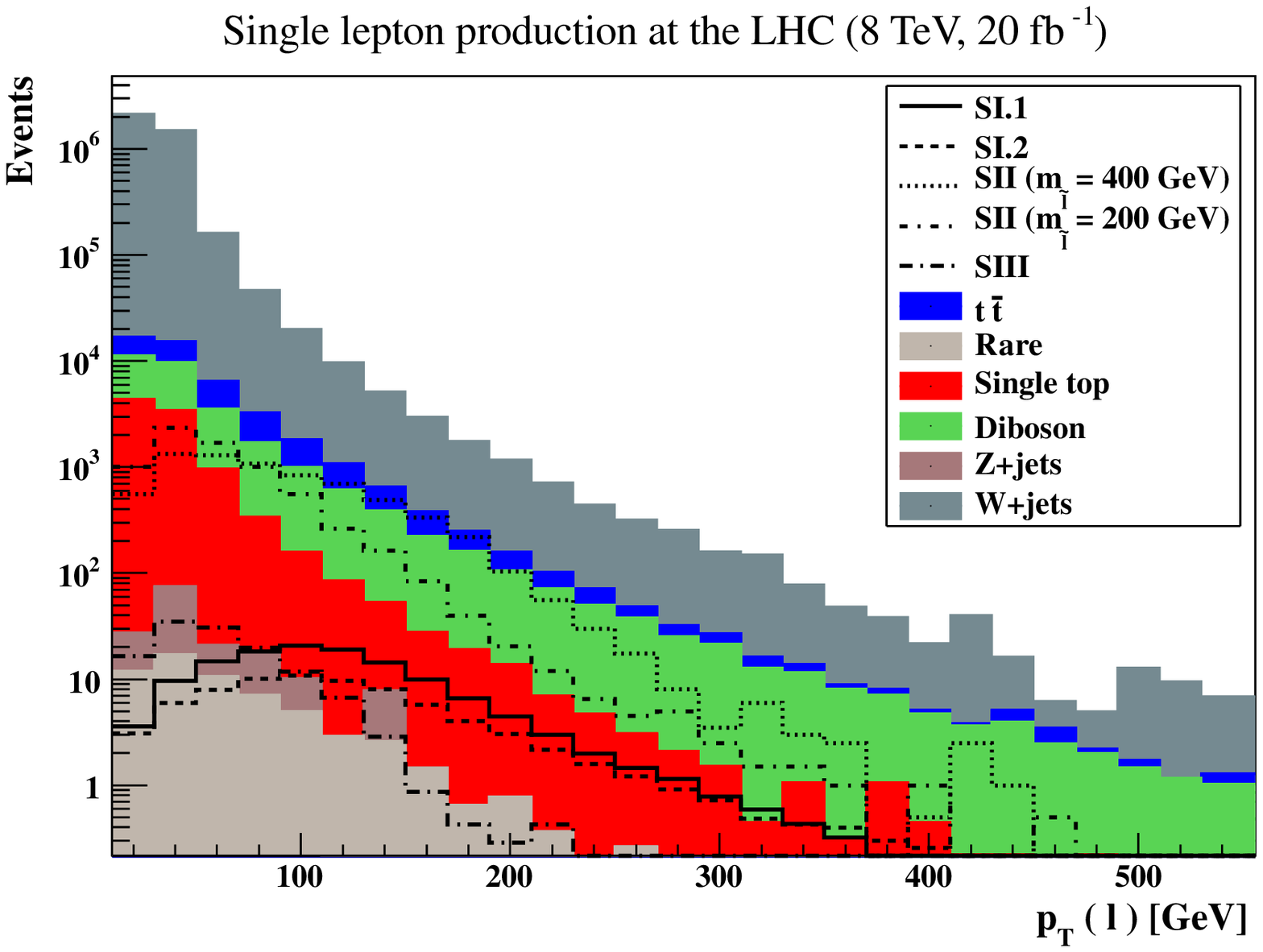}
 \end{center}
 \caption{\label{fig:monolep_pt1} Transverse-momentum spectrum of the identified lepton,
  after vetoing events with at least one
  $b$-tagged jet and selecting events with exactly one charged lepton, $M_T^W \geq 100$~GeV and
  $\slashed{E}_T \geq 100$~GeV.
  We considered 20 fb$^{-1}$ of LHC collisions at a center-of-mass energy of 8~TeV and present results for the
  different background contributions and for all the considered LRSUSY scenarios.}
\end{figure}

In order to reduce the SM contamination, we impose a constraint on the kinematical variable 
\be
  M_T^W =  \sqrt{ 2 p_T^\ell \slashed{E}_T \Big[ 1 - \cos \Delta
    \phi_{\ell,\slashed{E}_T}\Big] } \ , 
\label{eq:mtw}\ee
where $\Delta\phi_{\ell, \slashed{E}_T}$ stands for the angular distance, in the azimuthal direction with
respect to the beam, between the lepton and the missing energy. This variable
would be the $W$-boson transverse mass in the case where all the missing energy and the
identified lepton both originate from a $W$-boson decay. This quantity is expected to be smaller
in the context of the Standard Model, as illustrated on
Figure~\ref{fig:monolep_MT}, than in the LRSUSY case, which features cascade decays to multiple final state particles
contributing to the missing energy. In Figure~\ref{fig:monolep_MT}, we show results for the different contributions to the
SM background, upon which we superimpose signal distributions for the
LRSUSY scenarios designed in
Section~\ref{sec:benchmarks}. For the sake of simplicity, we fix the universal slepton masses to 400~GeV in all
cases but for the scenario {\bf SII}, where we consider both light and heavy sleptons with a mass
fixed either to 200~GeV or to 400~GeV.

On the basis of these results, we require the $W$-boson transverse mass
to satisfy $M_T^W~\geq~100$~GeV.
This reduces the background by a factor of 10, which is however
still dominated by $W$+jets events (99.6\%). Subdominant contributions include
diboson events and $t\bar t$ events
associated with topologies where possibly one or several leptons are not reconstructed.
In contrast,
$30\%-45\%$ (scenario {\bf SII}), $60\%-75\%$ (scenarios {\bf SI.1} and {\bf SI.2}) and
$70\%$ (scenario {\bf SIII}) of the signal events so far selected survive.

In Figure~\ref{fig:monolep_MET}, we present the other key observable of this single lepton
analysis, the missing transverse energy spectrum. One observes that the dominant background
contributions can be further suppressed by requiring large missing energy $\slashed{E}_T \geq 100$~GeV.
Once again, this selection does not affect the signal too much, 55\% to 99\% of the events
surviving in all scenarios, whereas the number of remaining background events
is reduced by a factor of six. The background consists still in 95.4\% of the cases of
$W$+jets events.
As shown in many experimental analyses, a combined selection
on the missing energy and on the $W$-boson transverse mass also allows
to keep the (non-simulated) multijet background under control (see, \eg, Ref.\
\cite{Aad:2012wm}), which justifies not considering them.

In Figure~\ref{fig:monolep_pt1}, we depict the lepton transverse-momentum distribution for the background
and the different signal scenarios. When large mass
splittings are present in LRSUSY spectra, gaugino-to-gaugino cascades induce very hard leptons as, \eg,
in scenarios {\bf SI.1}, {\bf SI.2} and {\bf SII}, in particular when sleptons are heavy. In this case, the
tails of the $p_T$ distributions even extend to values
greater than 200 GeV. These first two benchmarks however suffer from very small signal
cross sections, whereas the scenario {\bf SII} could lead to a promising discovery channel for LRSUSY
as gaugino mixing allows to produce new physics events at a larger rate. In contrast, for compressed
LRSUSY spectra such as in scenario
{\bf SIII}, we expect much softer leptons. The
lepton transverse momentum distribution indeed has its maximum at a $p_T$ value very close to the background one.
We optimized our selection focusing on the most promising cases and imposed $p_T(\ell) \geq 80$~GeV. This leads to a
good background rejection of a factor of about 3 together with a large signal efficiency of 50\%-70\%
in the relevant cases (and a smaller one for scenarios unlikely to be observed).

\begin{table}
  \begin{center}
    \begin{tabular}{c|c|c|c}
\hline
\hline
  Scenario     & Signal ($S$) & Background ($B$)& $S / \sqrt{S+B}$ \\
\hline
\hline  {\bf SI.1}                                   & $94.9 \pm 8.2$  & \multirow{5}{*}{$55332 \pm 247$}  & $0.40 \pm 0.08$\\
\cline{1-2}\cline{4-4}  {\bf SI.2}                   & $56.1 \pm 7.8$  &                                   & $0.24 \pm 0.07$\\
\cline{1-2}\cline{4-4}  {\bf SII} (200 GeV sleptons) & $1594 \pm 44$   &                                   & $6.68 \pm 0.36$\\
\cline{1-2}\cline{4-4}  {\bf SII} (400 GeV sleptons) & $3334 \pm 63$   &                                   & $13.8 \pm 0.5$\\
\cline{1-2}\cline{4-4}  {\bf SIII}                   & $31.8 \pm 6.2$  &                                   & $0.13 \pm 0.05$\\
\hline
\hline
    \end{tabular}
     \caption{\label{tab:cutflow1}
  Number of expected single-lepton events for 20 fb$^{-1}$ of LHC collisions at a center-of-mass energy of 8~TeV,
  given together with the associated statistical uncertainties, after applying
  all the selections described in the text. We present numbers of event $S$ for each of the signal scenarios introduced in
  Section~\ref{sec:benchmarks} after including a NLO $K$-factor set to 1.2 and for the background ($B$).
  The results are then converted in terms of LHC significance to LRSUSY signals in singly-leptonic final states.}
  \end{center}
\end{table}

After all selections, one finds that a very simple analysis based on a single lepton plus missing
energy topology is not suitable to probe most of the typical LRSUSY scenarios with light gauginos and
heavy higgsinos that can be built from low energy considerations. An exception is benchmark point
{\bf SII} featuring sensible gaugino mixings and enough mass splitting between the mass eigenstates
such that hard enough leptons are produced in their decays.
In this case, a sensitivity, defined as the ratio of the  number of selected signal events ($S$) to the squared
root of all selected signal ($S$) and background ($B$) events ($\sqrt{S+B}$), of
more than $5\sigma$ is expected for both {\bf SII} scenarios with 200~GeV and 400~GeV slepton
masses. In Table \ref{tab:cutflow1} we summarize
the results, expressed in terms of number of events surviving all selections and LHC sensitivity, for each of the
considered signal scenarios.

\subsubsection{A dileptonic signature}\label{sec:dil}
Based on the branching ratio tables of Section~\ref{sec:benchmarks}, dileptonic signatures are foreseen
to be quite frequent in the decay of a gaugino pair. They arise either from the dileptonic decay
of the first gaugino and a full hadronic or invisible decay of the second one, or from the singly-leptonic
decay of both superpartners. When accounting for geometrical acceptance ($|\eta(\ell)| \leq 2.5$),
transverse-momentum threshold ($p_T(\ell)\geq 10$~GeV) and isolation criteria (removal of leptons
too close to a jet), the signal efficiency of a $N_\ell=2$ charged lepton selection are
0.002 for scenario {\bf SIII}, where most of the decay products are too soft to be detected, to
about 20\%-30\% for the other scenarios.
Since gaugino-pair production rates are large for
scenarios of type {\bf SII} (see Table~\ref{tab:prod2}), these benchmarks are, as for the single lepton case, very promising for
observing hints of LRSUSY above the SM background. In this case, the latter overwhelms the LRSUSY {\bf SII} signal
by a factor of 300-500, depending on the slepton mass, this factor becoming $10^5-5 \times 10^5$ for all
the other scenarios.
After a dilepton selection, the background consists mainly of $Z$+jets events (99.5\%).

\begin{figure}
\begin{center}
  \includegraphics[width=.78\columnwidth]{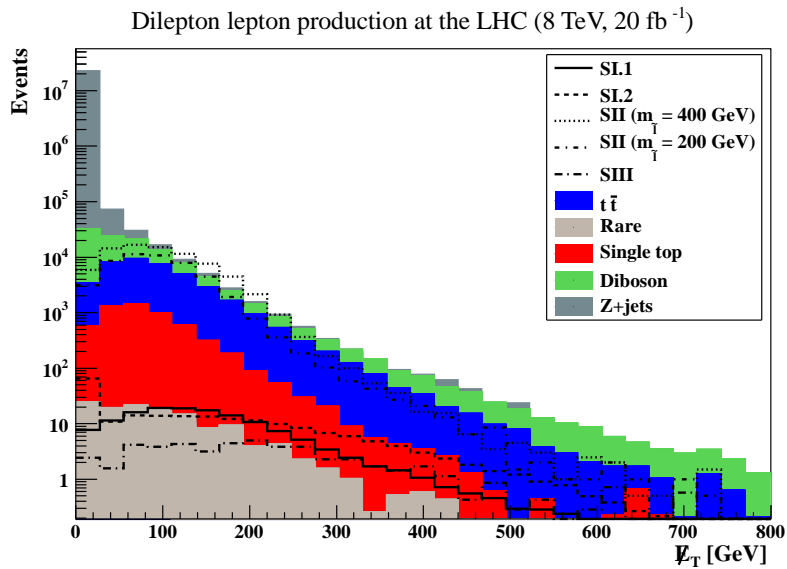}
 \end{center}
 \caption{\label{fig:dilep_MET} Missing transverse energy distribution after vetoing events with at least one
  $b$-tagged jet and selecting events with exactly two charged leptons.
  We considered 20 fb$^{-1}$ of LHC collisions at a center-of-mass energy of 8~TeV and present results for the
  different background contributions and for all the considered LRSUSY scenarios.}
\end{figure}

Consequently, it is tempting to impose a $Z$-veto on the invariant mass of the two leptons. However,
this selection has a very low signal efficiency so that we instead make the choice of requiring a combined
selection on the missing transverse energy and the transverse momentum of the leptons. In Figure~\ref{fig:dilep_MET},
we present the missing transverse energy spectrum of the different contributions to the SM background together
with the corresponding spectrum for the considered signal scenarios, \ie, all the four scenarios of Section~\ref{sec:benchmarks},
when the slepton mass is fixed to 400~GeV, and the scenario {\bf SII} in the case where the slepton mass is set to 200~GeV.
This leads us to  impose $\slashed{E}_T \geq 80$~GeV, which reduces the background by a factor
of about 575 and leads to the rejection of more than 99.9\% of the $Z$+jets events.
The surviving $Z$+jets events subsequently only contribute to 16\% of the SM background,
now dominated by $t\bar t$ events (45\%) and diboson events (33\%).

\begin{figure}
\begin{center}
  \includegraphics[width=.78\columnwidth]{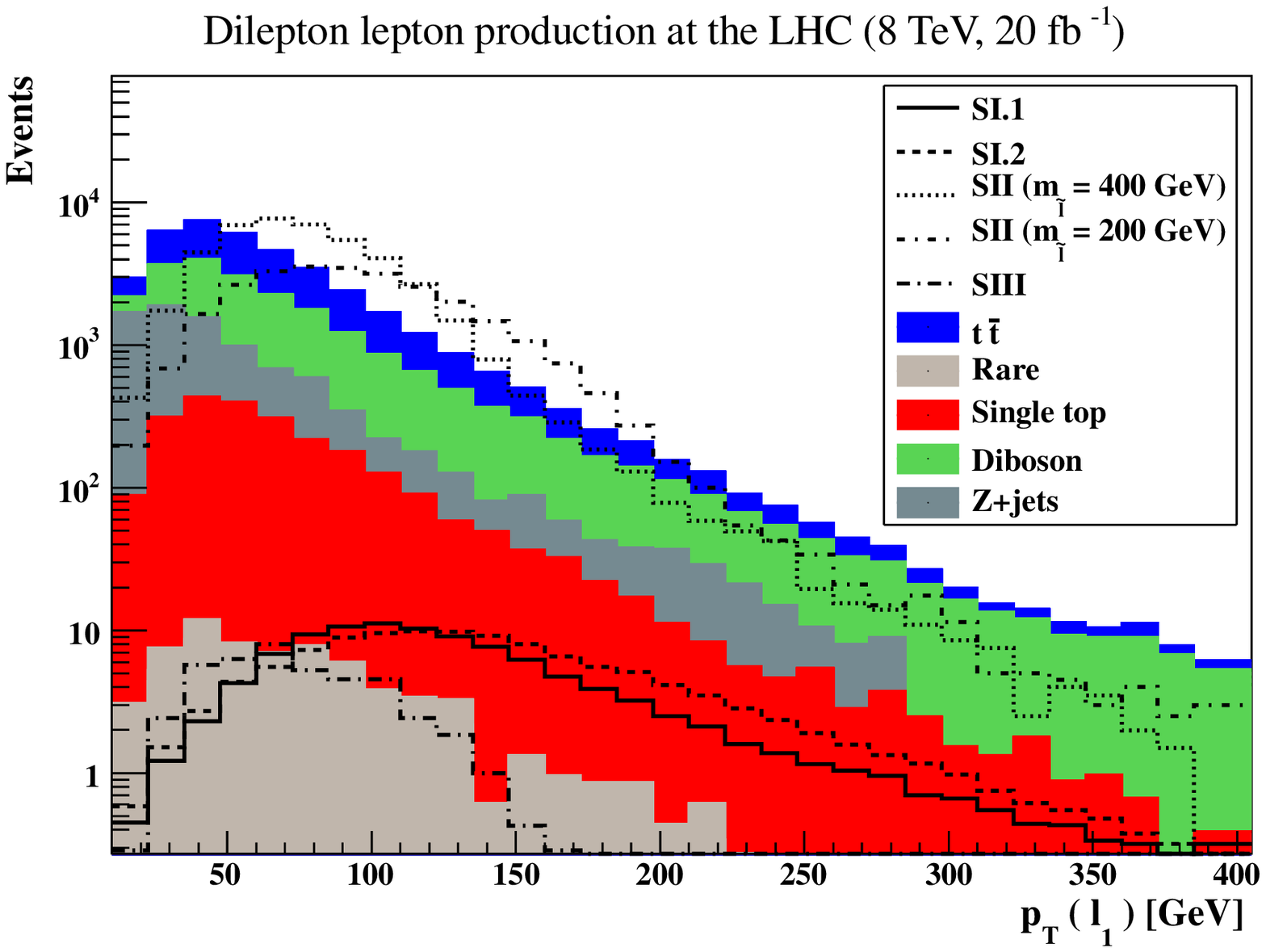}
 \end{center}
 \vspace{-.25cm}
 \caption{\label{fig:dilep_pt1} Transverse-momentum spectrum of the leading lepton $\ell_1$, after
  selecting events with exactly two charged leptons and no $b$-tagged jets, and at least 80~GeV of missing transverse energy.
  We considered 20 fb$^{-1}$ of LHC collisions at a center-of-mass energy of 8~TeV and present results for the
  different background contributions and for all the considered LRSUSY scenarios.}
\begin{center}
  \includegraphics[width=.78\columnwidth]{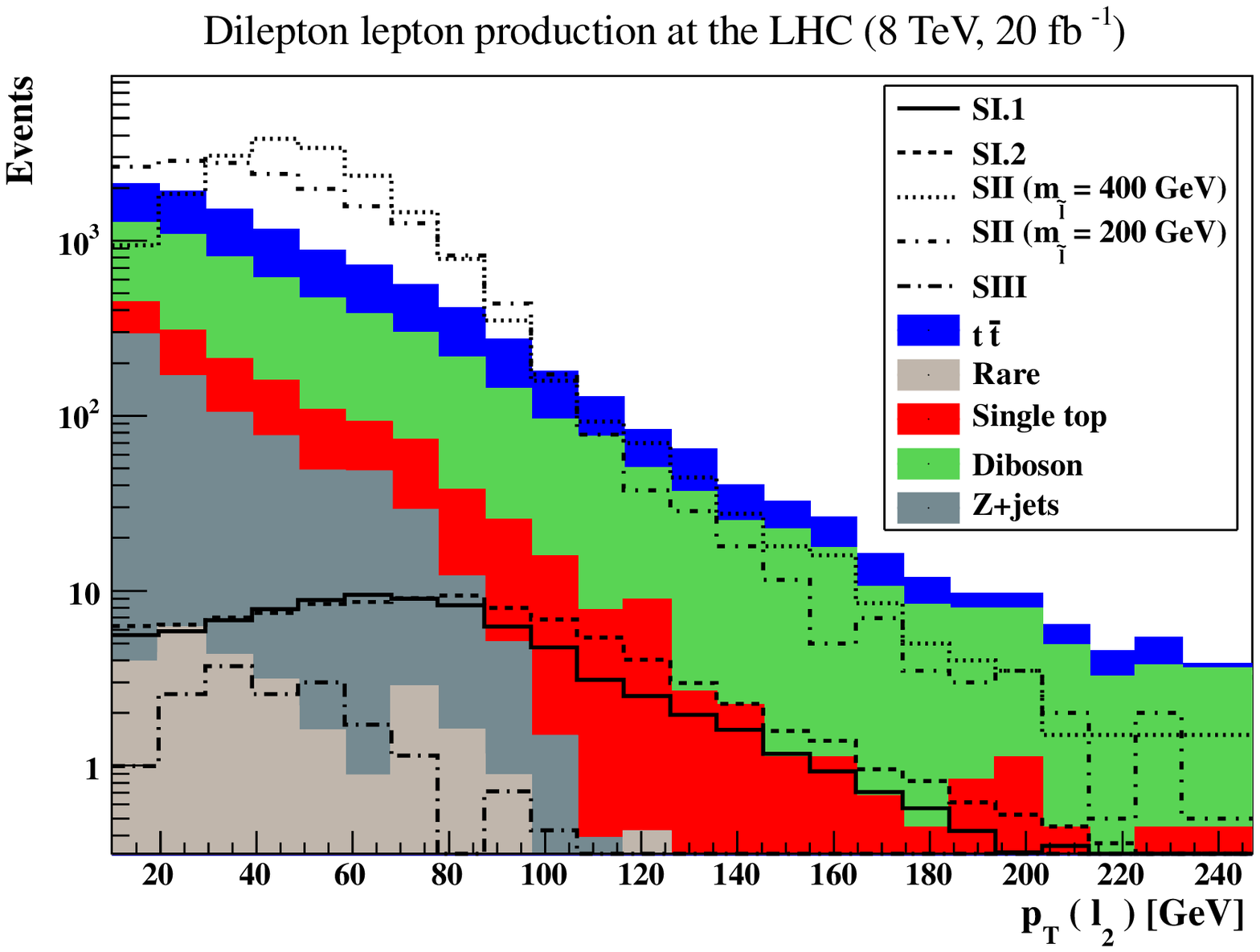}
 \end{center}
 \vspace{-.25cm}
 \caption{\label{fig:dilep_pt2} Transverse-momentum spectrum of the next-to-leading lepton $\ell_2$,
  after selecting events with exactly two charged leptons and no $b$-tagged jet, at least 80~GeV of missing transverse energy and
  a hard leading lepton with a $p_T$ greater than 80~GeV.
  We considered 20 fb$^{-1}$ of LHC collisions at a center-of-mass energy of 8~TeV and present results for the
  different background contributions and for all the considered LRSUSY scenarios.}
\end{figure}

In Figure \ref{fig:dilep_pt1}, we present the transverse-momentum distributions of the leading lepton for both
the remaining background and signal events. While scenarios of type {\bf SII} are already expected to be
observable, we further refine our analysis in order to try getting sensitivity to some of the other scenarios under
consideration. We hence include an additional restriction on the hardest of the two identified leptons $\ell_1$, requiring
its transverse momentum to satisfy $p_T (\ell_1) \geq 80$~GeV. In addition, we impose that the next-to-leading
lepton $\ell_2$ has to be hard, selecting events only if its $p_T$ is larger than 70~GeV. The effect
of this last restriction can be estimated from
Figure~\ref{fig:dilep_pt2} where we present the $p_T$ distribution of the second lepton
after all previous selections. Both these requirements ensure, together with our basic lepton isolation
criteria, that the non-simulated multijet background contributions
including fake leptons are under  control (see, \eg, Ref~\cite{Chatrchyan:2012oaa}).

\begin{table}
  \begin{center}
    \begin{tabular}{c|c|c|c}
\hline
\hline
  Scenario     & Signal ($S$) & Background ($B$)& $S / \sqrt{S+B}$ \\
\hline
\hline  {\bf SI.1}                                   & $41.2 \pm 6.8$  & \multirow{5}{*}{$1748.3 \pm 41.7$} & $0.97 \pm 0.32$\\
\cline{1-2}\cline{4-4}  {\bf SI.2}                   & $53.9 \pm 7.7$  &                                    & $1.27 \pm 0.36$\\
\cline{1-2}\cline{4-4}  {\bf SII} (200 GeV sleptons) & $2610 \pm 56$   &                                    & $39.5 \pm 1.2$\\
\cline{1-2}\cline{4-4}  {\bf SII} (400 GeV sleptons) & $2686 \pm 57$   &                                    & $40.3 \pm 1.2$\\
\cline{1-2}\cline{4-4}  {\bf SIII}                   & $2.6 \pm 1.8$   &                                    & $0.06 \pm 0.08$\\
\hline
\hline
    \end{tabular}
     \caption{\label{tab:cutflow2}
  Number of expected dilepton events for 20 fb$^{-1}$ of LHC collisions at a center-of-mass energy of 8~TeV,
  together with the associated statistical uncertainties, after applying
  all the selections described in the text. We present numbers of event $S$ for each of the signal scenarios introduced in
  Section~\ref{sec:benchmarks} after including a NLO $K$-factor set to 1.2 and for the background ($B$).
  The results are then converted in terms of LHC significance to LRSUSY signals in dileptonic final states.}
  \end{center}
\end{table}

The number of background events is subsequently found to be comparable with the
number of signal events in LRSUSY scenarios of class {\bf SII}, as shown in Table~\ref{tab:cutflow2}.
We also indicate in the table the expected significance for each benchmark point computed as the ratio of the number
of selected signal events to the squared root of the total number of predicted events. At this stage, background
consists mainly of top-antitop events
(44\%), diboson events (46\%) and single top events in the $tW$ channel (7\%).
Comparing with the single lepton analysis of
Section~\ref{sec:monol}, we  found that
the {\bf SII} scenarios are  likely to be observed with a very strong significance for both
chosen slepton mass. Unfortunately, there is still no sensitivity to the other considered scenarios.

\subsubsection{Signatures with three leptons or more}\label{sec:multil}
The two previous analyses are only sensitive to scenarios of class {\bf SII} mainly because they feature
a larger neutralino and chargino pair-production cross section due to the light associated masses.
When gauginos are  heavier, the LHC sensitivity to the corresponding LRSUSY signals is reduced,
as for our scenarios {\bf SI.1} and {\bf SI.2}, and it becomes difficult to extract the few LRSUSY signal events
from the overwhelming Standard Model background.
We therefore focus now on a multileptonic analysis requiring at least three charged leptons. This topology has the benefit  of a reduced Standard Model background so that new physics processes with  low cross sections can
possibly show hints in the observations.

\begin{figure}
\begin{center}
  \includegraphics[width=.78\columnwidth]{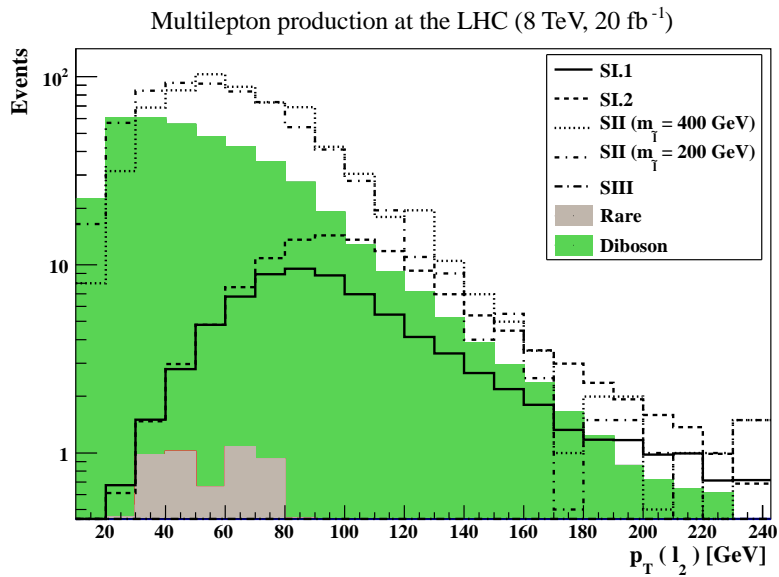}
 \end{center}
 \caption{\label{fig:multil_pt2} Transverse-momentum spectrum of the next-to-leading lepton $\ell_2$,
  after vetoing events with at least one
  $b$-tagged jet and selecting events with at least three charged leptons, at least 100~GeV of missing transverse energy and
  a hard leading lepton with a $p_T$ greater than 80~GeV.
  We considered 20 fb$^{-1}$ of LHC collisions at a center-of-mass energy of 8~TeV and present results for the
  different background contributions and for all the considered LRSUSY scenarios.}
\end{figure}

\begin{table}
  \begin{center}
    \begin{tabular}{c|c|c|c}
\hline
\hline
  Scenario     & Signal ($S$) & Background ($B$)& $S / \sqrt{S+B}$ \\
\hline
\hline  {\bf SI.1}                                   & $65.4 \pm 8.4$  & \multirow{5}{*}{$133.4 \pm 11.5$} & $4.64 \pm 1.03$\\
\cline{1-2}\cline{4-4}  {\bf SI.2}                   & $108  \pm 10$   &                                   & $6.98 \pm 1.09$\\
\cline{1-2}\cline{4-4}  {\bf SII} (200 GeV sleptons) & $259  \pm 18$   &                                   & $13.1 \pm 1.3$\\
\cline{1-2}\cline{4-4}  {\bf SII} (400 GeV sleptons) & $289  \pm 19$   &                                   & $14.1 \pm 1.3$\\
\cline{1-2}\cline{4-4}  {\bf SIII}                   & $\approx 0$     &                                   &  - \\
\hline
\hline
    \end{tabular}
     \caption{\label{tab:cutflow3}
  Number of expected multilepton events (with three ore more charged leptons)
  for 20~fb$^{-1}$ of LHC collisions at a center-of-mass energy of 8~TeV,
  together with the associated statistical uncertainties, after applying
  all the selections described in the text. We present numbers of event $S$ for each of the signal scenarios introduced in
  Section~\ref{sec:benchmarks} after including a NLO $K$-factor set to 1.2 and for the background ($B$).
  The results are then converted in terms of LHC significance to LRSUSY signals in multileptonic final states
  (with three or more charged leptons).}
  \end{center}
\end{table}

Signal efficiency for scenarios of class {\bf SI} is found to be moderate, reaching 20\%-30\%, in contrast
to the other scenarios for which it lies below 1\%. This low value is nevertheless compensated, in the case of scenario
{\bf SII}, by the large cross section. As  mentioned above, the Standard Model background is reduced (only about 5500 events are expected)
and mainly due to diboson
events (at 99.5\%). In the context of the Standard Model, the charged leptons included in
those events originate from a $Z$-boson or a $W$-boson leptonic decay. Therefore, we follow the same strategy
as in Section~\ref{sec:dil} and, instead of vetoing events with a lepton pair compatible with a $Z$-boson or imposing a selection
on the $W$-boson reconstructed transverse mass, we
require a selection based on the missing transverse energy and on the transverse momentum of the two
leading leptons. We hence impose that $\slashed{E}_T \geq 100$~GeV, together with the condition that the $p_T$ of the
two leading leptons is above thresholds of $p_T(\ell_1) \geq 80$~GeV and $p_T(\ell_2) \geq 70$~GeV. As shown in Figure~\ref{fig:multil_pt2}, where
we illustrate the last selection on the transverse momentum of the next-to-leading lepton $\ell_2$,
those simple cuts
are sufficient to highlight most of the considered LRSUSY signals from the diboson background.

The results are summarized in Table~\ref{tab:cutflow3} where we present, in addition to the Standard Model
expectation after all selections,
the number of signal events expected for each of the considered LRSUSY scenarios and the associated
significance given as $S/\sqrt{S+B}$. It can be seen that it reaches more than $3\sigma$ in all cases,
with the exception of scenario {\bf SIII} since its compressed spectrum does not allow for any visible signature.

\subsection{Comparison with the MSSM}
\label{sec:mssm}
We now turn to the comparison of LRSUSY signals with MSSM signals in the context of the analyses
introduced in Section~\ref{sec:lrsusy}. Assuming the observation of excesses in events with
a leptonic final state and a supersymmetric explanation for such excesses, we address the question of
probing the underlying theory and investigate if it exhibits more an MSSM or LRSUSY structure.
We first design MSSM scenarios with similar features
as the LRSUSY benchmarks of Section~\ref{sec:benchmarks}. To this aim, we follow the procedure below.
\begin{itemize}
  \item We start from a LRSUSY scenario and remove the neutralino and the chargino with the
    largest $SU(2)_R$ wino component.
  \item The two remaining LRSUSY neutralinos are identified as the two lighter neutralinos of the MSSM
    after neglecting their $SU(2)_R$ wino component. The masses are fixed to the same values in both models.
  \item The remaining LRSUSY chargino is identified as the lightest MSSM chargino after neglecting
    its possible $SU(2)_R$ wino component. Its mass is fixed to the same value in both models.
  \item We decouple all higgsinos in the MSSM.
  \item We then compute, by means of the {\sc FeynRules} \cite{Christensen:2008py,Duhr:2011se} and {\sc ASperGe} \cite{Alloul:2013fw}
    programs, the tree-level neutralino and chargino mass matrices and calculate the
    soft SUSY-breaking $U(1)_Y$ bino and $SU(2)_L$ wino
    mass parameters $M_1$ and $M_2$ leading to the proper mass eigenvalues.
\end{itemize}
This last step also enforces the choice for the mixing
parameters in the MSSM. We show the results in Table~\ref{tab:mssmscen}, giving the values
found for the $M_1$ and $M_2$ parameters. We hence design three scenarios
mimicking the LRSUSY scenarios {\bf SI.1}, {\bf SI.2} and {\bf SII}. Moreoer, we do not consider the LRSUSY
scenario {\bf SIII} as it is invisible at the LHC when considering leptonic final states.

\begin{table}
\begin{center}
\renewcommand{\arraystretch}{1.3}
\begin{tabular}{c | cc | ccc }
\hline
\hline
Scenario & $M_1$ [GeV] & $M_2$ [GeV] & $m_{\tilde\chi^0_1}$ [GeV] & $m_{\tilde\chi^0_2}$ [GeV] & $m_{\tilde\chi_1^+}$ [GeV]\\
\hline
{\bf SI.1} & 270 & 506 & 270&  500&  500 \\
\hline
{\bf SI.2} & 270 & 760 & 269&  747&  747 \\
\hline 
{\bf SII} & 112 & 254 & 111&  250&  250 \\
\hline \hline
\end{tabular}
\caption{\label{tab:mssmscen} MSSM scenarios {\bf SI.1}, {\bf SI.2} and {\bf SII} equivalent to their LRSUSY counterparts
of Section~\ref{sec:benchmarks}.}
\end{center} 
\renewcommand{\arraystretch}{1.0}
\end{table}

\begin{figure}
\begin{center}
  \includegraphics[width=.78\columnwidth]{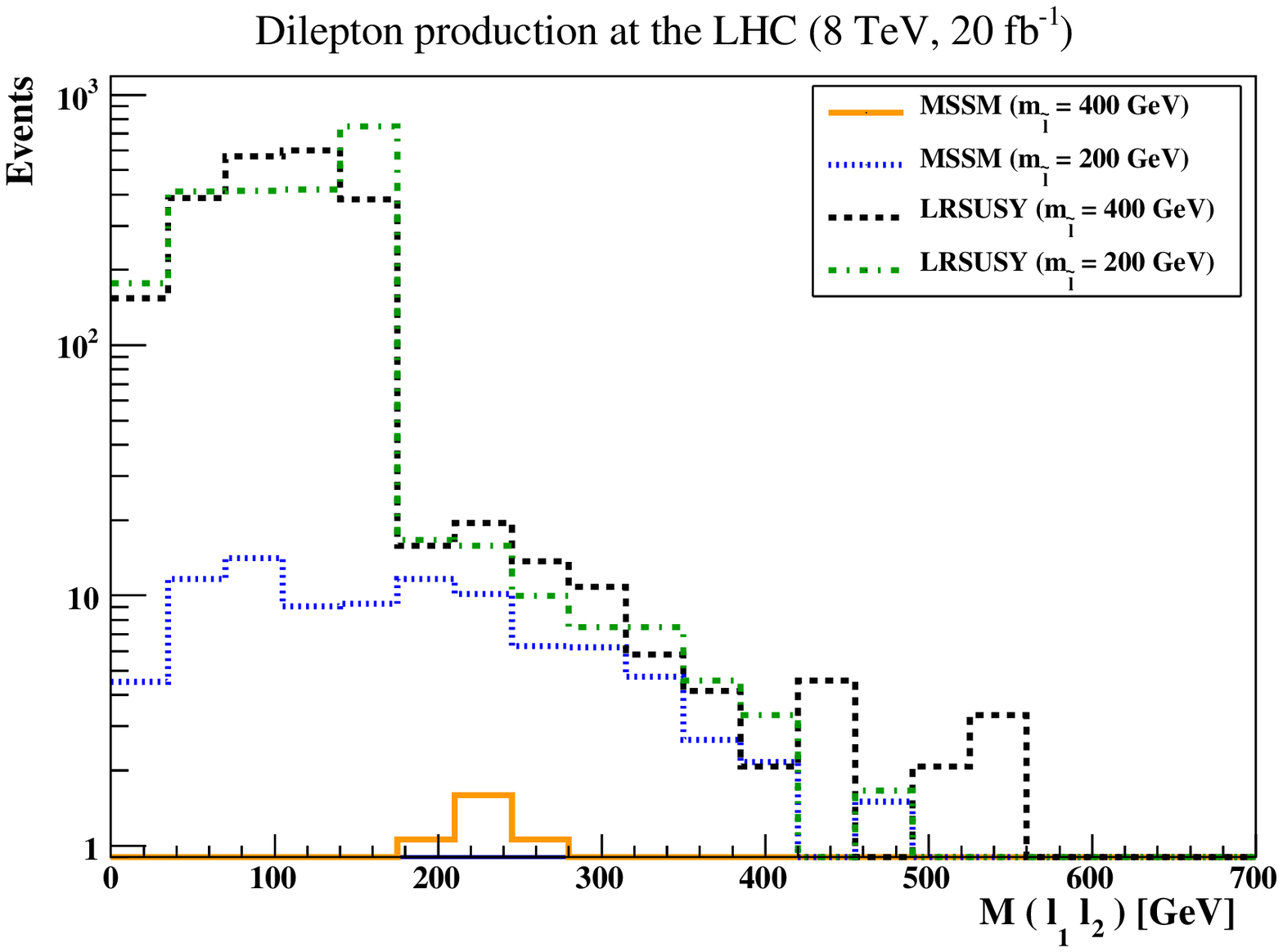}
  \includegraphics[width=.78\columnwidth]{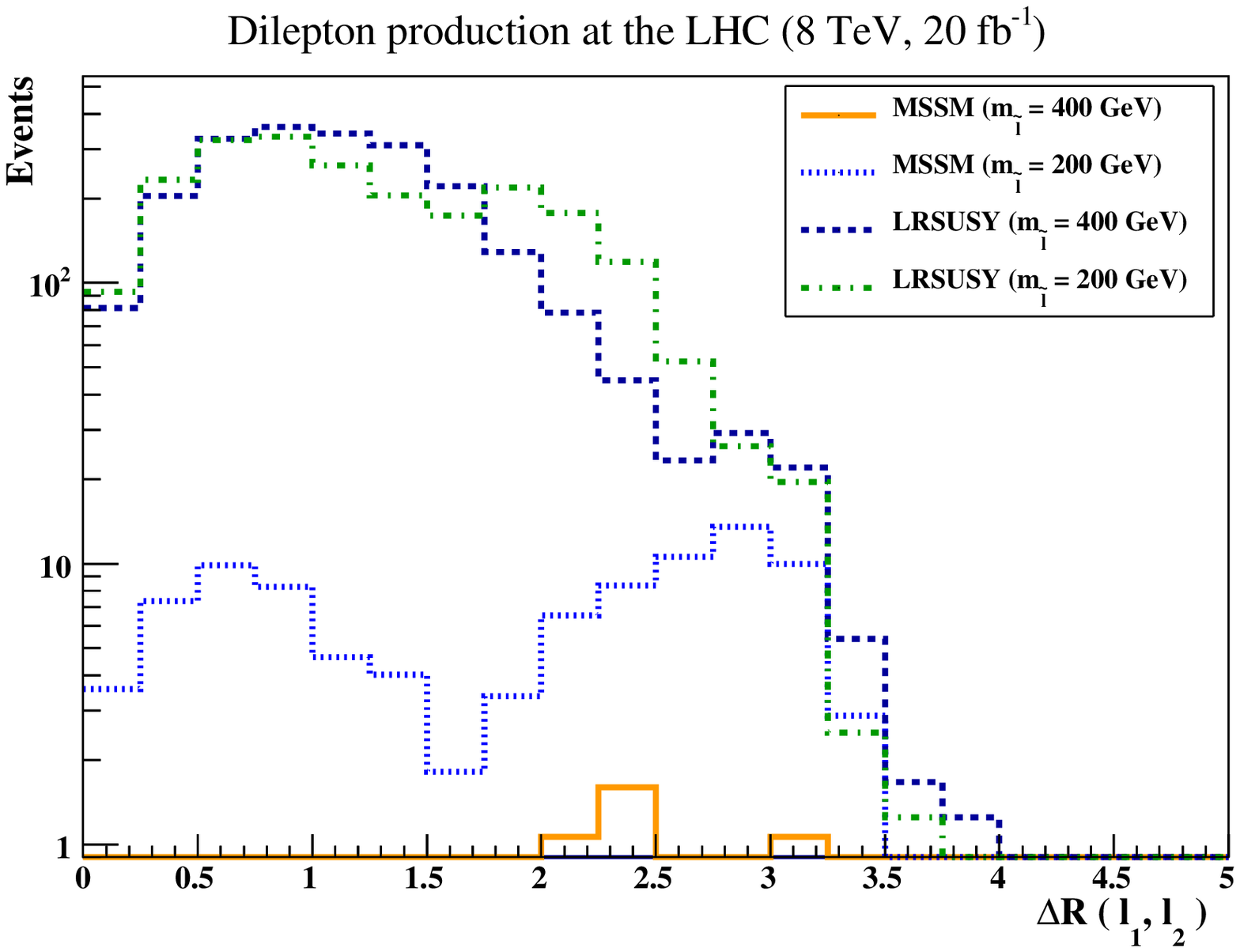}
\end{center}
\caption{\label{fig:dilep_mssm}Invariant mass (upper panel) and angular distance (lower panel) of a dilepton
 pair in MSSM and LRSUSY models after selecting events with exactly two charged leptons and no $b$-tagged jet,
 at least 80~GeV of missing
 transverse energy, and two leading leptons with a $p_T$ greater than 80~GeV and 70~GeV, respectively.
 We considered 20 fb$^{-1}$ of LHC collisions at a center-of-mass energy of 8 TeV and benchmark scenarios of type
 {\bf SII} with 200~GeV and 400~GeV sleptons.}
\end{figure}

For the sake of simplicity, we focus on the most promising channels, namely the
dilepton and multilepton (with three or more final state charged leptons) analyses. After
applying the selection criteria presented in Section~\ref{sec:dil} and Section~\ref{sec:multil}, we
investigate key distributions allowing to possibly disentangle a LRSUSY behavior from an MSSM one. Since in the dilepton case,
only  scenario {\bf SII} leads to a signal likely to be observed, we restrain the comparison to it and show the results
in Figure~\ref{fig:dilep_mssm}.  We present invariant mass and angular distance distributions among the two leading leptons for both
the LRSUSY and MSSM
cases, fixing the slepton masses either to 200~GeV or to 400~GeV. We observe that very few events are expected
in the case of the MSSM, in contrast to the LRSUSY one. Moreover, the shapes of the distributions are found also quite different,
so that they offer a possible way to distinguish both models assuming a given supersymmetric spectrum.

\begin{figure}
\begin{center}
  \includegraphics[width=.78\columnwidth]{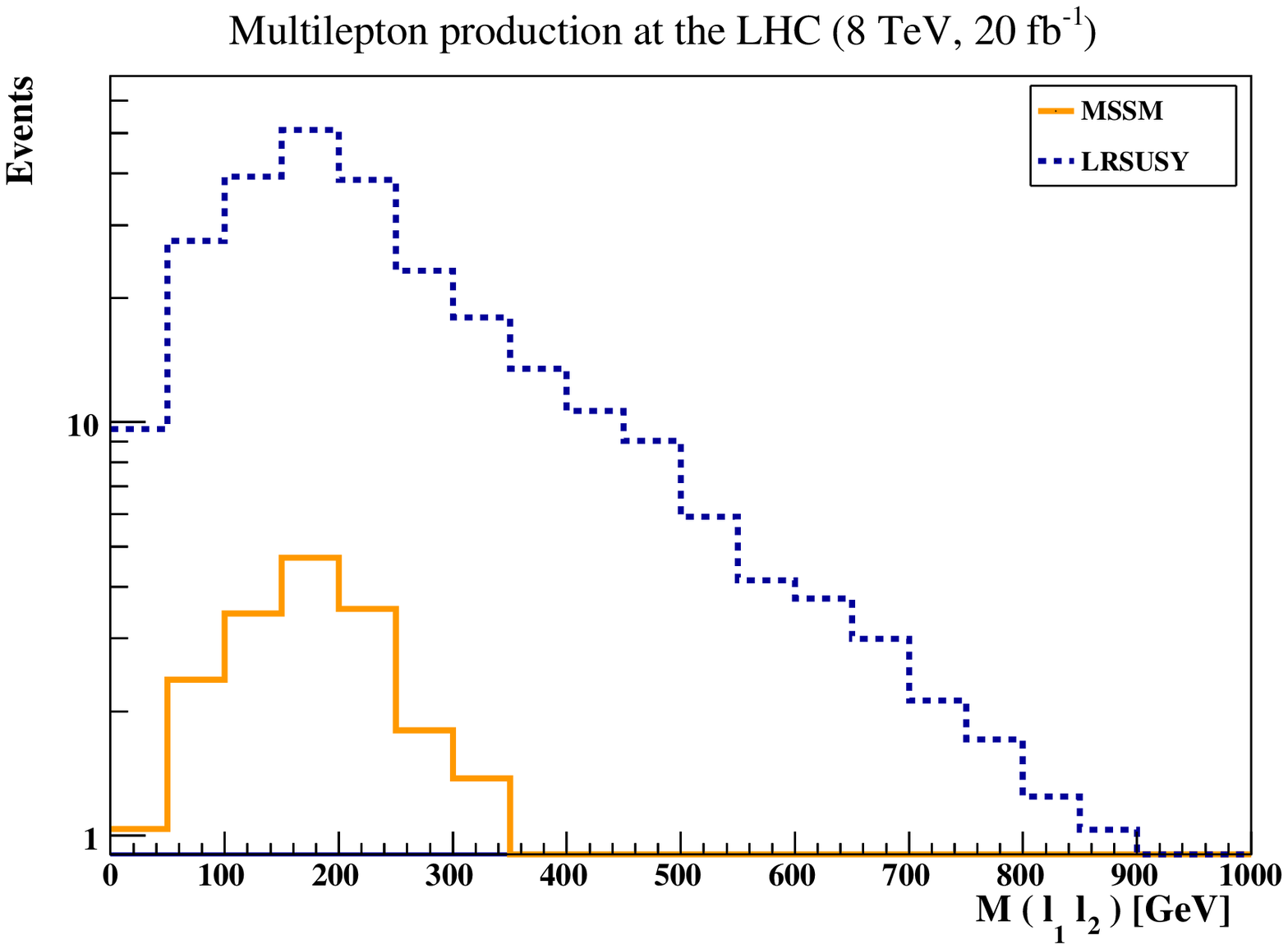}
  \includegraphics[width=.78\columnwidth]{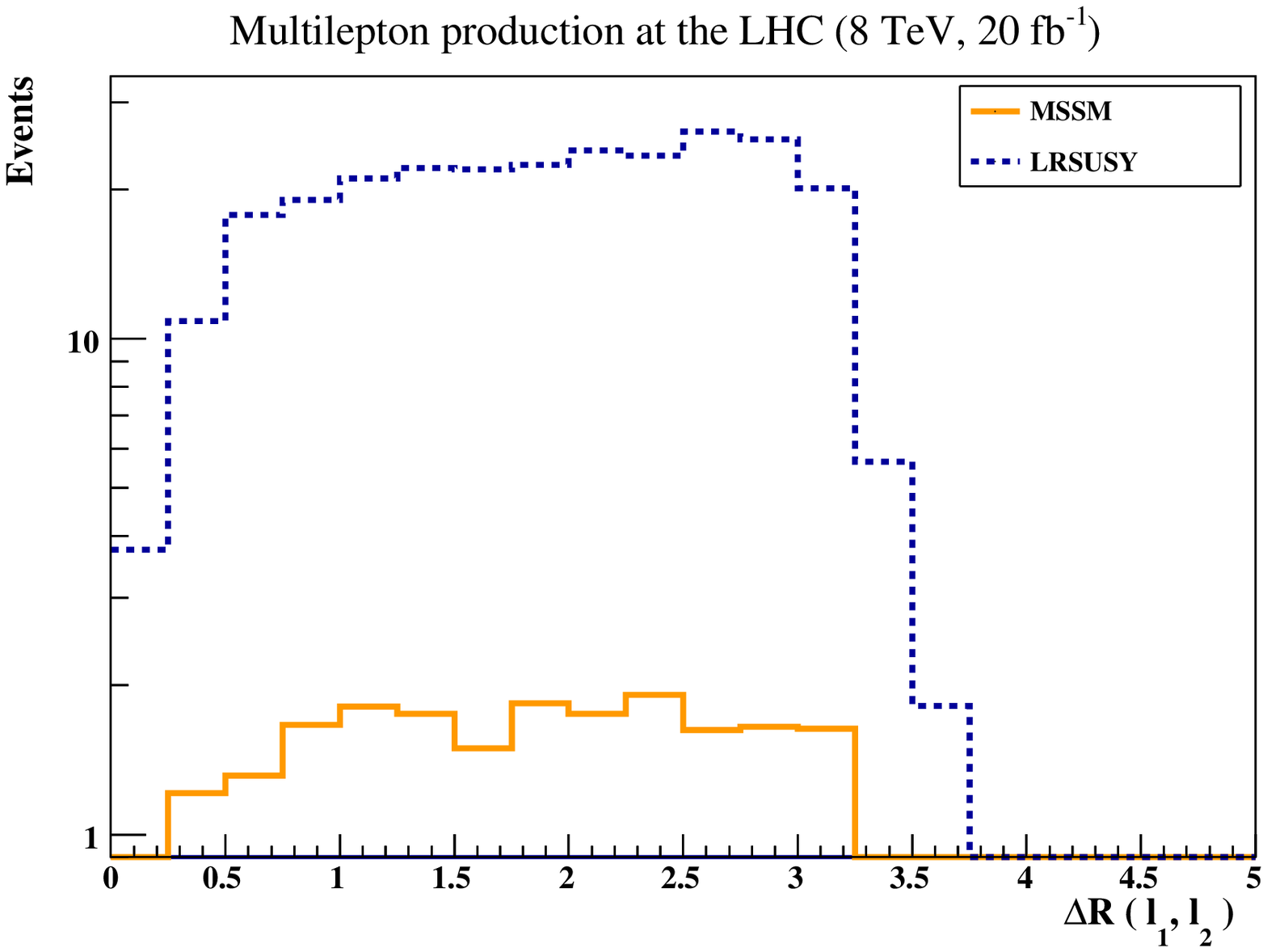}
\end{center}
\caption{\label{fig:multilep_mssm}Invariant mass (upper panel) and angular distance (lower panel) of a dilepton
 pair comprised of the two leading leptons $\ell_1$ and $\ell_2$
 in MSSM and LRSUSY models after selecting events with at least three charged leptons and no $b$-tagged jets,
 at least 100~GeV of missing
 transverse energy, and two leading leptons with a $p_T$ greater than 80~GeV and 70~GeV, respectively.
 We considered 20 fb$^{-1}$ of LHC collisions at a center-of-mass energy of 8 TeV and benchmark scenarios of type
 {\bf SI.1} with 400~GeV sleptons.}
\end{figure}

In the case of the multilepton analysis of Section~\ref{sec:multil}, no signal events are expected to survive
the selection strategy designed for
the MSSM counterparts of our scenarios {\bf SI.2} and {\bf SII}. In contrast, more than a $5\sigma$ sensitivity is expected
in the LRSUSY case. For scenarios of type {\bf SI.1}, lighter gaugino masses ensure that a few MSSM events
can be selected. We therefore illustrate their properties in Figure~\ref{fig:multilep_mssm}, where we present the invariant mass (upper panel)
and angular distance (lower panel) distributions of a particle pair comprised of the two leading charged leptons. As for
the dilepton case, a larger number of events is expected in LRSUSY models. However, the shapes of the distributions
are this time more similar. Nevertheless, if one restricts the spectra to their higher value bins containing many LRSUSY
events but very few MSSM events, this analysis again offers a possible
way to distinguish both cases.

\section{Conclusions}
\label{sec:conclusion}
In this work we explored the possibility that the associated production of charginos and neutralinos can be observed at the LHC.
We choose to work in a supersymmetric scenario where their production is likely to be enhanced, in a model
where the gauge symmetry is left-right symmetric, based on the $SU(3)_c \times SU(2)_L \times SU(2)_R \times U(1)_{B-L}$ group.
This model has twelve neutralinos (including two additional gauginos) and six (singly-charged) charginos (including an additional gaugino).
In comparison the MSSM has four neutralinos and two charginos. We present a complete description of the model, and follow
this by a choice of benchmark scenarios, chosen to highlight different mixing schemes and hierarchies among chargino and neutralino states.
After making some general observations about the patterns of chargino and neutralino decays in LRSUSY, and possible distinguishing signs
from similar decays in the MSSM, we present complete production and decay calculations for the benchmark scenarios,
classified according to the number of leptons in the final state. We proceed to event simulations, where we include the Standard Model backgrounds.
We devise methods to increase the signal to background significance, specifically for each one-lepton, two-lepton and three-or-more-lepton final states.
We complement our analysis by a simulation of events consistent with our benchmark scenarios in  the MSSM context (as much as possible).

Several general features emerge from our analysis. First, for most of the parameter space, with the exception of one scenario featuring large
gaugino mass splittings, the single lepton signal would not be visible at the LHC, as it is completely swamped by the background, even after
stringent  requirements on the missing transverse energy and transverse momentum of the lepton.
Imposing further  selection would then suppress both  signal  and background.  Second, two- and three-lepton signals are however visible above the background, especially in 
kinematical distributions associated with the leading and next-to-leading leptons.
Interesting, the most promising scenario is the one in which the LSP is a mixed state, a bino with a significant ${\tilde W}_R^0$ component, while the next-to-lightest
superpartner is pure left-handed wino.  This benchmark scenario raises above backgrounds after 
 applying the complete designed selection strategies, yielding  visible 
LRSUSY signal  at the LHC. And third, the number of events expected in LRSUSY scenarios of type {\bf SII} is
significantly above the expectations (one to two orders of magnitude, and different in shape) for the same events in a similar scenario in the MSSM
in the two-lepton final state,  but less so in the three-or-more-leptons, yielding a clear distinguishing signal from left-right supersymmetric models.

In a nutshell, enhanced production and decays of chargino and neutralino appear to be very promising signatures of supersymmetric models with extended gauge sectors, and in particular, for the left-right supersymmetric model, if these particles are light. These events complete favorably, and are complementary to, signals from the production and decays of doubly-charged higgsinos as means to test for left-right supersymmetry.

\acknowledgments
The authors are grateful to Alper Hayreter for enlightening discussions during early stages of this project
and to Eric Conte for invaluable help on its technical side.
The work of A.A. has been supported by a Ph.D.\ fellowship of the French ministry
for education and research and A.A., B.F. and M.R.T. have received partial support
from the Theory-LHC-France initiative of the
CNRS/IN2P3. The work of M.F. is supported in part by NSERC under grant number SAP105354.

\appendix
\label{sec:appendix}

\appendix
\label{sec:appendix}

\section{Conventions} \label{Sec:Model_Conv}
In this Section, we recall some basic features of Lie algebras in order to
fix the notations and correctly define the various relative signs which 
appear in the model described in Section~\ref{sec:model}.
Denoting $T_a$ the matrices of a unitary
representation ${\cal R}$ of a given Lie algebra $\mathfrak{g}$, it is well
known that the matrices $-T^t$, $-T^\star$ and $T^\dag$, \ie, the transposed, complex
conjugate and Hermitian conjugate matrices of $T$, span also representations
of the Lie algebra $\mathfrak{g}$\footnote{Let us note that for unitary
representations, $T = T^\dag$ and these matrices only span a single representation
of the corresponding Lie algebra.}. The representation spanned
by the matrices $-T_a^t$ is called the dual representation ${\cal R}^\ast$, the
one spanned by the matrices $-T_a^\star$ the complex conjugate representation
$\overline{{\cal R}}$ and the one spanned by the matrices $T^\dag_a$ the dual of
the complex conjugate representation $\overline{{\cal R}}^*$.

However, it may happen that some of these four representations are isomorphic.
As an example, for $SU(2)$, if we denote $\utilde{\bf 2}$ the two-dimensional
(fundamental) representation spanned by the Pauli matrices $\frac12 \sigma_i$,
it turns out that we get the isomorphism $\utilde{\bf 2} \cong \utilde{\bf 2}^*
\cong \utilde{\bf \bar 2}$, since
\be
    -\sigma^t_i = \sigma_2\ \sigma_i\ \sigma_2^{-1} 
  \quad \text{and} \quad
    -\sigma^\star_i = \sigma_2\ \sigma_i\ \sigma_2^{-1} \ .
\ee
The first of these two isomorphisms allows to raise or lower the two-dimensional
indices by the mean of the invariant $SU(2)$ tensors $\e_{ij}$
 and $\e^{ij}$ defined by
$\e_{12}=-\e_{21}=-\e^{12}=\e^{21}=1$. Hence, for a field $\psi^i$ lying in
the $\utilde{\bf 2}$ representation,
\be \label{eq:su2}
  \psi_i = \e_{ij}\ \psi^j \quad \text{and} \quad  \psi^i = \e^{ij}\ \psi_j  \ .
\ee
In the case of $SU(3)$, we  similarly get, for the three-dimensional
representations, the relation $\utilde{\bf 3}^* \cong \utilde{\bf \bar 3}$.

In this paper, we denote by $i$ and $i^\prime$ typical indices of the
two-dimensional representation of $SU(2)_L$ and $SU(2)_R$, respectively, while
we associate to the three-dimensional representation of $SU(3)_c$ 
indices  labeled by $m$.

\section{Gauge boson mass matrices}\label{app:ewsb}

We can extract the gauge boson mass matrices from the Higgs field kinetic terms, 
\be  \bsp
  M^2_{V^0} = &\ \bpm 
     \frac14 g_L^2 \big(4 v_L^2 \!+\! v^2 \!+\! v^{\prime 2}\big) &
    -\frac14 g_L g_R \big(v^2+v^{\prime 2}\big) &
    - {\hat g}g_L v_L^2 \\
    -\frac14 g_L g_R \big(v^2+v^{\prime 2}\big) &
     \frac14 g_R^2     \big(4 v_R^2 \!+\! v^2 \!+\! v^{\prime 2}\big) &
    - {\hat g}g_R v_R^2 \\
    -{\hat g} g_L v_L^2 &
    -{\hat g} g_R v_R^2 &
     {\hat g}^2 \big(v_L^2 + v_R^2\big) \\ 
   \epm \ , \\ 
  M^2_{V^\pm} = &\ \bpm 
     \frac14 g_L^2 \big(2 v_L^2 \!+\! v^2 \!+\! v^{\prime 2}\big) &
    -\frac12 g_L g_R (vv^\prime)^\ast \\
    -\frac12 g_L g_R (vv^\prime) &
     \frac14 g_R^2 \big(2 v_R^2 \!+\! v^2 \!+\! v^{\prime 2}\big) \\
   \epm \ , 
\esp \ee
where we have introduced the abbreviations
\be\bsp
&\ v_L^2 = v_{1L}^2+v_{2L}^2\ , \quad
  v_R^2 = v_{1R}^2+v_{2R}^2\ , \\ &\
  v^2 = v_1^2+v_2^2 \ , \quad
  v^{\prime 2} = v_1^{\prime2} + v_2^{\prime 2} \ , \quad
  vv^\prime = v_1 v_1^\prime e^{i \alpha_1} + v_2 v_2^\prime e^{i \alpha_2} \ .
\esp\label{eq:reducedvevs} \ee
In the limit of the vev hierarchy of Eq.\ \eqref{eq:vevhier}, the mass matrix of
the neutral gauge bosons, usually diagonalized with the help of an orthogonal
$3\times 3$ matrix $U_g^0$, is  diagonalized through two independent
rotations of angles $\theta_W$ and $\phi$. This follows  the model
breaking pattern. After the breaking of the $SU(2)_R\times U(1)_{B-L}$, the neutral $W_{R\mu}^3$ and
$\hat B_\mu$ fields mix to a massless state, which will be identified to the
hypercharge field $B'_\mu$, and a massive $\Zp$-boson, which will
decouple from the breakdown process. When the electroweak symmetry is eventually
broken at a lower scale to electromagnetism, the hypercharge field and the neutral
$W_{L\mu}^3$ field  then mix to a massless state identified to the photon
$A_\mu$ and to a massive state, the $Z$-boson. The mixing matrix takes  a
simple form,
\be\bsp
   \bpm Z_\mu \\A_\mu \\  \Zp_\mu \epm = &\
   \bpm 
     \cw & -\sw\sin\phi & -\sw\cos\phi\\
     \sw & \cw\sin\phi & \cw\cos\phi\\
     0&\cos\phi & -\sin\phi\\
   \epm
   \bpm W_{L\mu}^3 \\ W_{R\mu}^3 \\ \hat B_\mu \epm \\  = &\
   \bpm 
     \frac{e}{g_Y} & -\frac{e\ g_Y}{g_L\ g_R} & -\frac{e\ g_Y}{g_L\ {\hat g}}\\
     \frac{e}{g_L} & \frac{e}{g_R} & \frac{e}{{\hat g}}\\
     0&\frac{g_Y}{{\hat g}} & -\frac{g_Y}{g_R}\\
   \epm
   \bpm W_{L\mu}^3 \\ W_{R\mu}^3 \\ \hat B_\mu \epm  \ ,
\esp\label{eq:neutralmixing}\ee
where in the last line, we have expressed the mixing angles as functions of the
electromagnetic coupling constant $e$, the hypercharge coupling constant $g_Y$
and the unbroken gauge group coupling constants $g_L$, $g_R$ and ${\hat g}$.
The physical masses are given by
\be \bsp
   m_\Zp^2 =&\  v_R^2 ({\hat g}^2 + g_R^2) = \frac{g_R^2}{\cos\phi^2} v_R^2\ , \\
   m_Z^2   =&\ \frac14 \Big[ g_L^2 + \sin\phi^2 g_R^2\Big] 
    v^2  = \frac{g_L^2}{4 \cos^2\theta_W} 
    v^2   \ ,
\esp\ee
and the photon stay massless. We have also the following relations, linking the
mixing angles to the coupling constants,
\be\bsp
 &\  \cos\phi=\frac{g_R}{\sqrt{g_R^2+{\hat g}^2}}\ , \quad
   \sin\phi=\frac{{\hat g}}{\sqrt{g_R^2+{\hat g}^2}}, \quad\\
 &\  \cw=\frac{g_L}{\sqrt{g_R^2 \sin\phi^2 + g_L^2}}\ , \quad 
   \sw=\frac{g_R \sin\phi}{\sqrt{g_R^2 \sin\phi^2 + g_L^2}} \ .
\esp\ee

Turning to the charged sector, the mass matrix is usually diagonalized by a
$2\times 2$ unitary matrix $U_g^\pm$. However, in the approximation of Eq.\
\eqref{eq:vevhier}, $U_g^\pm$ tends to the identity matrix, the mass of the two
eigenstates being simply 
\be
  m_W^2 = \frac{g_L^2}{4}  v^2 
  \quad \text{and} \quad
  m_\Wp^2 = \frac12 g_R^2 v_R^2 \ .
\ee

\bibliographystyle{JHEP}

\end{document}